\documentclass[reprint,amsmath,amssymb,floatfix,onecolumn,superscriptaddress,aps,pre]{revtex4-2}
\usepackage{hyperref}
\usepackage{graphicx}
\usepackage{dcolumn}
\usepackage{threeparttable}
\usepackage{bm}
\usepackage{color}
\usepackage{xcolor}
\usepackage{dsfont}
\usepackage{ulem}
\hypersetup{
    colorlinks=true,
    linkcolor=blue,
    citecolor=blue,   
   urlcolor=blue,
   }
\def\bea{\begin{eqnarray}}
\def\eea{\end{eqnarray}}
\def\beq{\begin{equation}}
\def\eeq{\end{equation}}

\def\D{\Delta}

\def\t{\tau}

\def\Re{{\rm Re}}

\def\s{\sigma}

\def\la{\langle}
\def\ra{\rangle}
\def\nn{\nonumber}

\def\d{\delta}

\def\w{\omega}
\def\g{\gamma}

\def\d{\delta}
 
\def\la{\langle}
\def\ra{\rangle}

\def\g{\gamma}

\begin{document}

\title[Inertial harmonic chain]{Crossover dynamics and non-Gaussian fluctuations in inertial active chains}

\author{Manish Patel}%
\email[]{manish.patel@iopb.res.in}
\affiliation{Institute of Physics, Sachivalaya Marg, Bhubaneswar-751005, Odisha, India}
\affiliation{Homi Bhabha National Institute, Anushakti Nagar, Mumbai 400094, India}

\author{Subhajit Paul}%
\email[]{spaul@physics.du.ac.in}
\affiliation{Department of Physics and Astrophysics, University of Delhi, Delhi-110007, India}

\author{Debasish Chaudhuri}
\email[Corresponding author:~]{debc@iopb.res.in}
\affiliation{Institute of Physics, Sachivalaya Marg, Bhubaneswar-751005, Odisha, India}
\affiliation{Homi Bhabha National Institute, Anushakti Nagar, Mumbai 400094, India}
\date{\today}%

\begin{abstract}
We study the dynamics of inertial active particles in a one-dimensional chain with harmonic nearest-neighbor interactions, highlighting the interplay of persistence, interaction, and inertial timescales. Using a Green’s function approach, we derive the mean-squared displacement (MSD) and mean-squared change in velocity (MSCV), revealing multiple crossovers between ballistic, diffusive, and subdiffusive regimes and providing analytic expressions for scaling coefficients and crossover times. Non-Gaussian deviations in active Brownian particles are captured through excess kurtosis, reflecting heavy-tailed, finite-support, or bimodal distributions that evolve systematically over time. Time-dependent probability distributions exhibit distinct data collapses within different temporal regimes, confirming the robustness of the scaling behavior. Overall, this framework connects multiparticle interactions to microscopic dynamics, revealing experimentally accessible signatures of inertia in active matter.

\end{abstract}

\maketitle
\section{Introduction}
Active matter systems comprise self-propelled agents that consume energy to generate directed motion, giving rise to a rich spectrum of nonequilibrium phenomena at both individual and collective levels~\cite{Bechinger2016,Marchetti2013, Romanczuk2012, Ramaswamy2019, Reimann2002}. 
These systems violate the equilibrium fluctuation-dissipation relation and remain perpetually out of equilibrium. 
Examples span a wide range of scales, from microscopic swimmers such as motor proteins, bacteria, and cells to macroscopic collectives such as flocks of birds, schools of fish, and herds of animals. 
Synthetic realizations, including self-propelled colloids and Janus particles, mimic these natural systems and have enabled precise experimental control over active dynamics. 
Theoretical descriptions of such self-propelled motion are commonly provided by three minimal models: (i) run-and-tumble particles (RTPs)~\cite{Berg_Brown_1972, Schnitzer_1993}, (ii) active Brownian particles (ABPs)~\cite{Howse2007, Romanczuk2012, Cates2013}, and (iii) active Ornstein-Uhlenbeck particles (AOUPs)~\cite{Szamel_14,Fodor16}. 
Although their microscopic rules differ, all three share exponentially decaying correlations in active noise and yield equivalent second-moment dynamics.

In many biological and synthetic settings, active processes take place in crowded environments where interparticle interactions play a crucial role. These interactions give rise to collective phenomena such as velocity-alignment-induced flocking~\cite{Vicsek95}, swarming~\cite{Marchetti2013},  and motility-induced phase separation (MIPS)~\cite{tailleur_08, Fily2012}.
While these behaviors are well established numerically, analytical understanding of interacting active systems remains limited. 
Most theoretical efforts have employed coarse-grained hydrodynamic frameworks~\cite{Marchetti2013, Cates15},  which provide insight into collective instabilities but rarely capture microscopic time-scale competition in interacting active particles. To move beyond coarse-grained descriptions, we use a minimal, analytically tractable model—a one-dimensional harmonic chain of active particles—that captures the interplay of inertia, activity, and interactions at the microscopic level.

Traditionally, active-particle dynamics have been studied in the overdamped limit, an excellent approximation for microscopic colloids whose inertial relaxation times ($\sim 100$~ns) are orders of magnitude smaller than their persistence times ($\sim 1$ -- $10$~s)~\cite{kurzthaler2018}. 
However, for larger active entities -- such as vibrobots, hexbugs, camphor boats, or self-propelled granular rods -- the inertial relaxation time becomes comparable to or exceeds the persistence time, making inertia an essential component of the dynamics~\cite{Scholz2018, Leoni20, Dauchot19, Tapia-Ignacio_2021, Devereux2021, Mukundarajan2016, Rabault2019}. 
Recent studies have shown that inertia profoundly influences both asymptotic and steady-state properties~\cite{Scholz2018, Mandal2019, Lowen_2020, Patel_2023, Fazelzadeh23, Patel_2024, Dutta24, Karan24, Khali24, Dutta25}, modifying velocity correlations~\cite{Caprini21}, entropy production~\cite{Shankar18, Patel_2024}, and even stabilizing active-nematic phases~\cite{Chatterjee21}. 
Departures from equilibrium in active systems diminish at large inertia~\cite{Caprini22, Patel_2024}.  

In one dimension, interparticle interactions impose strong constraints on motion, giving rise to single-file diffusion (SFD) -- a universal subdiffusive behavior known from passive Brownian systems~\cite{Harris1965, Kollmann_2003, Lutz_2004}. 
Recent analytical and numerical studies have extended this phenomenon to active systems~\cite{SlowmanEvansBlythe2017, DasDharKundu2020, LeDoussalMajumdarSchehr2019, SlowmanEvansBlythe2016, KourbaneHousseneErignouxBodineauTailleur2018, DolaiDasKunduDasguptaDharKumar2020, BanerjeeJackCates2022, AgranovRoKafriLecomte2023}, demonstrating how persistence and self-propulsion modify the SFD amplitude~\cite{Paul24, SinghKundu2021}. 
Spectral analyses of inertial active chains driven by colored noise~\cite{Marconi2024} have provided a mode-resolved view of activity-induced collective excitations, yet existing treatments remain limited to linear dynamics and cannot capture tracer-level statistics, dynamical crossovers, or non-Gaussian fluctuations.
A comprehensive analytical treatment that unifies these regimes -- linking inertial and overdamped dynamics, collective and tracer-level observables, and Gaussian to non-Gaussian statistics -- has thus far been lacking.

Here, we develop such a framework to investigate the dynamics of an inertial chain of active particles coupled through harmonic interactions in one dimension. Using an analytical Green's function approach, we derive key two-point observables -- mean-squared change in velocity (MSCV), and mean-squared displacement (MSD) -- spanning all dynamical regimes governed by three intrinsic timescales: inertial ($\tau_m$), persistence ($\tau_a$), and interaction ($\tau_k$).

Our analysis reveals six distinct intermediate time regimes, characterized by multiple crossovers from ballistic to diffusive and, eventually, to subdiffusive behavior. The corresponding crossover times are derived explicitly, and the late-time dynamics are shown to exhibit well-defined steady-state features in the velocity domain, including an effective kinetic temperature. The second-moment scaling properties hold across all three representative models of active particles -- ABP, RTP, and AOUP. Theoretical predictions are validated through numerical simulations. 

Extending beyond second moments, we computed the excess kurtosis and full velocity and displacement distributions for ABPs, revealing non-Gaussian features -- including sign reversals of kurtosis and bimodal, finite-support, or heavy-tailed distributions -- that distinguish ABPs from AOUPs. The time-dependent probability distributions exhibit distinct data collapses within different temporal regimes, confirming the robustness of the scaling behavior.

This framework connects the effects of multiparticle interactions to microscopic dynamics, uncovering signatures of inertia in active matter that are experimentally measurable and potentially relevant for systems such as active granular particles and active solids.

The remainder of this paper is organized as follows. 
Section~\ref{sec:model} introduces the model and the theoretical framework. 
Section~\ref{sec:second} presents results for second-moment observables, including the velocity autocorrelation, MSCV, and MSD. 
Section~\ref{sec:fourth} discusses higher-order statistics and excess kurtosis, while Section~\ref{sec:prob} analyzes the dynamical scaling of probability distributions. 
Finally we conclude in Section~\ref{sec:conclusion}.

\section{Model and methodology} \label{sec:model}

We consider a chain of $N$ inertial active particles with mass $m_p$ interacting via harmonic forces with a coupling strength $k$ to its nearest neighbor. 
The displacement around lattice positions and velocity of $l^{\mathrm th}$ particle is represented by ${x_l}$  and $v_{l}(=\dot{x}_l)$ for $l = 1, 2, ..., N$. 
The underdamped equation of motion for the system can be written as
\begin{align}
\label{eq:xdot}
        \dot{X} &= V \nn\\
    m_p \dot{V}+\gamma V &=-\Phi X + \gamma V^a,
\end{align}
 where $X(t) = (x_1, x_2,...,x_N)^T$, $V(t) = (v_1, v_2,...,v_N)^T$, $V^a(t) = (v^a_1,v^a_2,...,v^a_N)^T$ and the $\Phi$ is a symmetric tri-diagonal matrix with elements $\Phi_{l,m}=k(2\delta_{l,m}-\delta_{l,m-1}-\delta_{l,m+1})$. The active velocities obey 
 \bea
 \langle v^a_l (t_1) v^a_m (t_2)\rangle = v_0^2 \,\delta_{l,m}\, e^{-|t_1-t_2|/\tau_a}
 \label{eq_va_corr}
 \eea
 with persistence time $\t_a$. The dynamics involves inertial relaxation time $\t_m = m_p/\g$ and bond relaxation time $\t_k = \g/k$, with $\g$ denoting viscous drag. 
 
 {\it Mapping ABP, RTP, and AOUP:}
 Despite their distinct microscopic dynamics, the ABP, RTP, and AOUP models are equivalent at the level of second-order statistics, provided their parameters are matched via $v_0 = v_{\rm abp}/\sqrt{2} = v_{\rm rtp} = v_{\rm ou}$ and $\tau_a = \tau_{\rm abp} = \tau_{\rm rtp} = \tau_{\rm ou}$  (see Appendix-\ref{sec:mapping}). 
 Each model generates zero-mean active velocities with exponentially decaying autocorrelations obeying Eq.\eqref{eq_va_corr}. However, differences emerge beyond second moments: AOUP yields Gaussian fluctuations, whereas ABP and RTP exhibit inherently non-Gaussian dynamics.

{\it Numerical integration:} The equations of motion can be integrated using the velocity Verlet scheme, using the unit of time $\t_a$ and length $\s$, the bond length. Activity is controlled by ${\rm Pe}=v_0/\t_a \s$. See Appendix-\ref{app_verlet} for methodological details. Periodic boundary conditions are used unless specified otherwise.

 {\it Green's function formulation:}
 To compute correlation functions for a chain of particles driven by active forces, we employ the temporal Fourier transforms: 
 $\tilde{X}_\omega =  (1/2 \pi) \int_{-\infty}^{\infty} dt \,X(t) e^{-i \omega t}$ and $\tilde{V}_\omega =  (1/2 \pi) \int_{-\infty}^{\infty} dt \,V(t) e^{-i \omega t}$. 
 Applying the equations of motion and incorporating active noise in the frequency domain leads to (for $\tilde{V}_{\omega}$ see Appendix \ref{appendix:vomega}): 
\bea
\label{eq:xtilde_omega}
    \Tilde{X}_{\omega} = \gamma  \Tilde{\mathcal{G}}(\omega)\Tilde{V}^a(\omega), ~~
    \tilde{V}_{\omega} = i \gamma  \omega \Tilde{\mathcal{G}}(\omega)\Tilde{V}^a(\omega),
\eea
with the Green's function 
\bea
\Tilde{\mathcal{G}}(\omega)= (\Phi+i\gamma \omega \mathds{I}-m\omega^2 \mathds{I})^{-1}.
\eea
Here, $\Tilde{V}^a(\omega)$ denotes the Fourier-transformed active velocities. This formalism elegantly captures the system's response to active forcing through the matrix $\Tilde{\mathcal{G}}(\omega)$. 
We use this to compute key observables. The two-time velocity correlation matrix is given by:
\bea
\label{eq:generic_corrv}C^{v} (t) = \la V (t) V^T(0) \ra .
\eea
Using the known correlation of active noise in Fourier space, 
$\la \tilde v^a_l (\omega_1) \tilde v^a_m(\omega_2)\rangle = v_0^2\, [2 \pi \tau_a (2 - i \tau_a( \omega_1 + \omega_2)) \delta(\omega_1 + \omega_2) \delta_{l,m}]/(1 - i \omega_1 \tau_a)(1 - i \omega_2 \tau_a)$,  we obtain the explicit component form: 
\bea
C^{v}_{l,m} (t) = \frac{\gamma^2 v_0^2 \t_a}{\pi} \int_{-\infty}^{\infty} \frac{ \omega^2 \sum_{k} \tilde{\cal G}_{l,k}(\omega) \tilde{{\cal G}}^{T}_{m,k} (-\omega)}{1 + \omega^2 \t_a^2} e^{i \omega t} d\omega. \nn\\
\label{eq:corrv_comp}
\eea
Analogously, the displacement correlation matrix is:
\bea
\label{eq:generic_corr}{C}^{x} (t) &&= \la X (t) X^T(0) \ra,
\eea
giving
\bea
\label{eq:corr_x}
    {{C}}^{x}_{l,m} (t) &&=\frac{\gamma^2 v_0^2 \t_a}{\pi} \int_{-\infty}^{\infty} \frac{ \sum_{k} \tilde{\cal G}_{l,k}(\omega) \tilde{{\cal G}}^{T}_{m,k} (-\omega)}{1 + \omega^2 \t_a^2} e^{i \omega t} d\omega.
\eea 
These correlation functions provide a complete statistical description of the velocity and displacement fields of particles in the chain.

\section{Second moments} \label{sec:second}
In this section, we examine the second moment of velocity and position. We examine the two-time autocorrelation and equal-time spatial correlation in velocity. 
Further, we analyze dynamical crossovers in MSCV and MSD, using analytic results and their comparison against numerical simulations.

\subsection{Dynamical Space Time Correlations}

To characterize the temporal dynamics of particle velocities, we consider the system-averaged two-time velocity autocorrelation function:
$
{ C}_{vv} (t) = (1/N) \sum_{l = 1}^N C^v_{l,l} (t).
$
Using Eq.~(\ref{eq:corrv_comp}), this can be expressed as:  
\bea
C_{vv}(t) =  {\rm Re} \int_{-\infty}^\infty d\w\, \tilde C_{vv}(\w) e^{i\w t} 
\label{eq:autocorr}
\eea
with $\tilde C_{vv}(\w)=[\mathcal{B}_v(\omega)/ (1 +\omega^2 \tau_a^2)]\, [{\gamma^2 v_0^2 \t_a}/{\pi}\,]$ 
where the spectral weight  $\mathcal{B}_v(\omega)$ encapsulates the system's dynamical response:
\bea \label{eq:bvomega}
\mathcal{B}_v(\omega)=\frac{\omega^2}{N} \sum_{l,m} \tilde{\mathcal{G}}_{l,m}(\omega) \tilde{\mathcal{G}}_{m,l}(-\omega) 
 =\frac{1}{\gamma^2} {\mathrm{Re}} \left [  \frac{\sqrt{\omega \tau_k}}{\sqrt{( \omega \tau_m-i)(4 - \omega \tau_k (\omega \tau_m - i))}}\right].
\eea
A detailed derivation of ${\cal B}_v(\omega)$ is provided in Appendix~\ref{appendix:bvomega_bxomega}.

\subsection{Dynamical crossovers and scaling regimes}
In this section, we analyze the time evolution of the mean-squared change in velocity and the mean-squared displacement to identify distinct scaling regimes and the dynamical crossovers between them. 

\begin{figure*}
    \centering
    \includegraphics[width=1\linewidth]{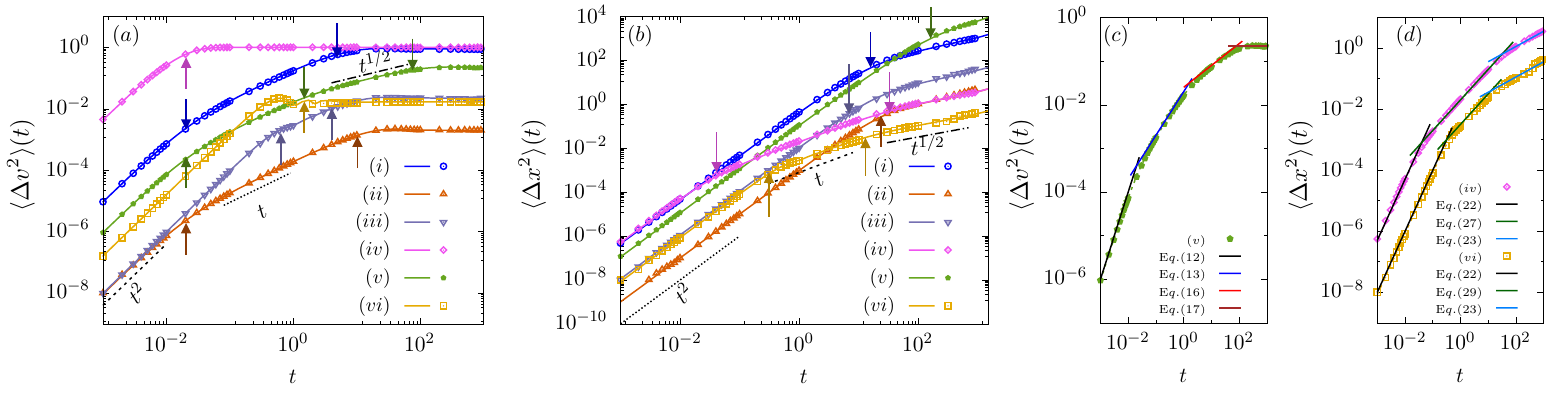}
    \caption{
Time evolution of MSCV ($a,\,c$) and MSD ($b,\, d$) for different parameter sets $(\tau_m, \tau_k, \tau_a)$: ($i$) (0.01, 10, 10), ($ii$) (10, 100, 0.01), ($iii$) (10, 0.01, 10), ($iv$) (0.01, 100, 0.01), ($v$) (0.01, 5, 100), and ($vi$) (10, 0.01, 0.1). 
Solid lines show numerical integrations of Eqs.~(\ref{bulk_msv_full_form}) and (\ref{bulk_msd_full_form}), respectively, for MSCV and MSD, and points denote simulations at fixed activity $v_0 = 1$. 
Arrows indicate crossover times, and panels ($c,d$) illustrate corresponding dynamical scalings.
}
    \label{fig:msd_msv}
\end{figure*}

\subsubsection{Mean-squared change in velocity}

We characterize the velocity dynamics through the mean-squared change in velocity (MSCV), defined as
\[
{\cal C}^v_{l,m}(t)=\langle [v_l(t)-v_l(0)][v_m(t)-v_m(0)]\rangle,
\]
which, for $l=m$, gives the MSCV of the $l$-th particle. Averaging over all particles yields
\bea
\langle {\D v}^2 \rangle(t)=\frac{ \gamma^2 v_0^2 \tau_a}{\pi}\int_{-\infty}^{\infty}
\mathcal{B}_v(\omega) \,
\frac{2[1-\cos(\omega t)]}{1+\omega^2\tau_a^2}\,d\omega,
\label{bulk_msv_full_form} 
\eea
where $\mathcal{B}_v(\omega)$ encodes the frequency-dependent response of the chain. Although an exact analytical solution is not available, numerical evaluation of this integral shows excellent agreement with simulations (Fig.~\ref{fig:msd_msv}$a$). In the following, we outline how to obtain analytical insights in the intermediate time properties and crossovers between them.

{\bf Early-time regime: }
At short times ($t \ll \tau_a, \tau_k$), the integral in Eq.~(\ref{bulk_msv_full_form}) is dominated by large frequencies, yielding
\begin{equation}
\label{eq:msv_ballistic}
\langle {\D v}^2 \rangle(t) = \frac{v_0^2 \tau_a}{ \pi} \int_{-\infty}^{\infty} 
\frac{4\sin^2 (\omega t/2)}{ {(\omega^2 \tau_m^2 +1)(1 +\omega^2 \tau_a^2)}}\, d\omega 
\approx \frac{v_0^2 }{\tau_m(\tau_a+ \tau_m)} t^2 + \mathcal{O}(t^4).
\end{equation}
This corresponds to ballistic scaling ($\langle {\D v}^2 \rangle \sim t^2$), as shown in Fig.~\ref{fig:msd_msv}($a$). 
The ballistic coefficient decreases with both inertia and persistence time.

{\bf Intermediate-time regimes:}
Depending on the relative magnitudes of the intrinsic timescales—namely inertia ($\tau_m$), persistence ($\tau_a$), and interaction ($\tau_k$)—the MSCV exhibits up to six intermediate temporal regimes: $(i) \t_m \ll t \ll (\t_a, \t_k)$, $(ii) \t_a \ll t \ll  (\t_m, \t_k)$, $(iii) \t_k \ll t \ll (\t_a, \t_m)$, $(iv) (\t_m,\t_a) \ll t \ll \t_k$, $(v) (\t_m,\t_k) \ll t \ll \t_a$ and $(vi) (\t_k,\t_a) \ll t \ll \t_m$.
Figure~\ref{fig:msd_msv}($a$) illustrates the corresponding dynamical behaviors.

{\bf($i$)} $\tau_m \ll t \ll (\tau_a, \tau_k)$:
In this regime, ${\cal B}_v(\omega)$ can be approximated as ${\cal B}_v \simeq 1/\gamma^2$, and $(1 + \omega^2 \tau_a^2) \simeq \omega^2 \tau_a^2$. Substituting these into Eq.~\eqref{bulk_msv_full_form} gives
\begin{equation}
\label{eq:mscv1}
\langle {\D v}^2 \rangle(t) = \frac{v_0^2 \tau_a}{\pi} \int_{-\infty}^{\infty} \frac{4 \sin^2(\omega t/2)}{\omega^2 \tau_a^2}\, d\omega \approx \frac{2 v_0^2}{\tau_a} t,
\end{equation}
indicating a diffusive behavior of MSCV at intermediate times, with an effective diffusion coefficient $D_v = v_0^2/\tau_a$.
The crossover from the early ballistic regime to this diffusive regime occurs at $t_1^{c} = 2\tau_m(\tau_a + \tau_m)/\tau_a$, obtained by equating Eq.~\eqref{eq:msv_ballistic} with Eq.~\eqref{eq:mscv1}.
For $\tau_a \neq \tau_k$, MSCV can further exhibit a crossover from diffusive to subdiffusive behavior at later times (discussed in case~($v$)).
In the absence of such a crossover, MSCV saturates at long times, with the diffusive-to-saturation crossover occurring at $t_2^{c} = \tau_a^2 \sqrt{\tau_k}/\sqrt{(\tau_a + \tau_m)(4\tau_a^2 + \tau_a \tau_k + \tau_k \tau_m)}$, obtained by matching Eq.~\eqref{eq:mscv1} to the steady-state value in Eq.~\eqref{eq:ss_mscv}.
The corresponding behavior is shown by line~($i$) in Fig.~\ref{fig:msd_msv}($a$), for parameters $(\tau_m, \tau_k, \tau_a) = (0.01, 10, 10)$, where MSCV transitions from ballistic to diffusive motion at $t_1^{c} = 0.02$, followed by saturation at $t_2^{c} = 4.47$.

{\bf($ii$)} $\t_a \ll t \ll (\t_m, \t_k) $: In this limit, ${\cal B}_v (\omega)$ simplifies as $ {\cal B}_v (\omega) \approx \frac{1}{\gamma^2} \Re \left[\frac{1}{i (\omega \t_m - i)} \right] \approx \frac{1}{\gamma^2 (\omega^2 \t_m^2  + 1)} \approx \frac{1}{\gamma^2 \omega^2 \t_m^2}$ while $(1 + \omega^2 \t_a^2) \approx 1$. Substituting these forms yields 
\bea
\langle {\D v}^2 \rangle(t) = \frac{ v_0^2 \t_a }{ \pi } \int_{-\infty}^{\infty} \frac{4 \sin^2(\omega t/2)}{  \omega^2 \t_m^2} d\omega \approx \frac{2 v_0^2 \t_a}{ \t_m^2}t.
\eea
MSCV thus exhibits diffusive behavior with an inertia-dependent diffusion coefficient $D_{v} = v_0^2 \t_a/\t_m^2$. The ballistic-to-diffusive crossover occurs at $t_{1}^c = 2 \t_a (\t_a + \t_m)/\t_m$, obtained from comparing the above equation with Eq.~(\ref{eq:msv_ballistic}). Moreover, MSCV saturates at  $t_{2}^c = \t_m^2 \sqrt{\t_k}/\sqrt{(\t_a + \t_m)(4 \t_a^2 + \t_a \t_k + \t_k \t_m)} $, obtained from equating Eqs.~\eqref{eq:msv_ballistic} and \eqref{eq:ss_mscv}.
For parameters $(\tau_m, \tau_k, \tau_a) = (10, 100, 0.01)$, MSCV transitions from ballistic to diffusive motion at $t_1^{c} = 0.02$ and saturates at $t_2^{c} = 9.99$, as shown by line~($ii$) in Fig.~\ref{fig:msd_msv}($a$).

{\bf($iii$)} $\tau_k \ll t \ll (\tau_m, \tau_a)$:
In this limit, ${\cal B}_v(\omega) \simeq \sqrt{\tau_k/(4\tau_m \gamma^4)}$ and $(1 + \omega^2 \tau_a^2) \simeq 1$, leading to
\begin{equation}
\langle {\D v}^2 \rangle(t) = \frac{v_0^2 \tau_a}{2\pi } \sqrt{\frac{\tau_k}{\tau_m}} \int_{-\infty}^{\infty} \frac{4 \sin^2(\omega t/2)}{\omega^2 \tau_a^2}\, d\omega \approx \frac{v_0^2}{\tau_a} \sqrt{\frac{\tau_k}{\tau_m}}\, t.
\end{equation}
MSCV therefore exhibits diffusive behavior with a diffusion coefficient $D_v = \tfrac{v_0^2}{2\tau_a} \sqrt{\tfrac{\tau_k}{\tau_m}}$, dependent on all three timescales.
The ballistic-to-diffusive crossover occurs at $t_1^{c} = \sqrt{\tau_m \tau_k}\,(\tau_a + \tau_m)/\tau_a$, obtained from Eq.~\eqref{eq:msv_ballistic}.
For $\tau_a \neq \tau_k$, MSCV may further transition to a subdiffusive regime before saturation (discussed in case~($v$)).
In the absence of this crossover, saturation occurs at $t_2^{c} = 2\tau_a^2 \sqrt{\tau_m}/\sqrt{(\tau_a + \tau_m)(4\tau_a^2 + \tau_a \tau_k + \tau_k \tau_m)}$ obtained from comparing with Eq.\eqref{eq:ss_mscv}.
For $(\tau_m, \tau_k, \tau_a) = (10, 0.01, 10)$, MSCV transitions from ballistic to diffusive motion at $t_1^{c} = 0.63$ and saturates at $t_2^{c} = 3.53$, as shown by line~($iii$) in Fig.~\ref{fig:msd_msv}($a$).

{\bf($iv$)} $( \t_m, \t_a)  \ll t \ll \t_k $: In this regime, ${\cal B}_v(\omega) \simeq 1/\gamma^2$ and $(1 + \omega^2 \tau_a^2) \simeq 1$, indicating that MSCV has effectively reached its late-time saturation value, given in Eq.~\eqref{eq:ss_mscv}. For $\tau_m \neq \tau_a$, an early-time ballistic-to-diffusive crossover may occur, as in cases~($i$) or ($ii$). In the absence of such a crossover, MSCV transitions directly from ballistic to saturation.
The crossover time is obtained by comparing the ballistic regime, Eq.~\eqref{eq:msv_ballistic}, with the saturation value, Eq.~\eqref{eq:ss_mscv}, yielding
$t_{1}^c = ([2 \t_a \t_m \sqrt{\t_k(\t_a + \t_m)} \,]/\sqrt{4 \t_a^2 + \t_a \t_k + \t_k \t_m} \,)^{1/2}$.
For $(\tau_m, \tau_k, \tau_a) = (0.01, 100, 0.01)$, MSCV exhibits a direct crossover from ballistic to saturation at $t_1^c = 0.014$, as shown by line~(iv) in Fig.~\ref{fig:msd_msv}($a$).

{\bf ($v$) $( \t_m, \t_k)  \ll t \ll \t_a $ }: 
In this regime, $B_v(\omega) \simeq \sqrt{\omega \tau_k}/(2^{3/2} \gamma^2)$ and $(1 + \omega^2 \tau_a^2) \simeq \omega^2 \tau_a^2$, giving
\begin{equation}
\langle {\D v}^2 \rangle(t) = \frac{\sqrt{2 \tau_k} v_0^2}{\pi \tau_a} \int_{-\infty}^{\infty} \frac{\sin^2(\omega t/2)}{\omega^{3/2}} d\omega
\simeq \frac{2 v_0^2}{\tau_a} \sqrt{\frac{\tau_k}{\pi}}\, t^{1/2}.
\end{equation}
MSCV therefore exhibits sub-diffusive $t^{1/2}$ scaling with a coefficient $2 v_0^2 \sqrt{\tau_k}/(\tau_a \sqrt{\pi})$.
The crossover from ballistic or diffusive to sub-diffusive behavior can be identified by comparing the corresponding MSCV with the above equation.
The line plot ($v$) in Fig.~\ref{fig:msd_msv}($a$) shows the MSCV evolution for $(\tau_m, \tau_k, \tau_a) = (0.01, 5, 100)$, exhibiting a crossover from ballistic to diffusive to sub-diffusive dynamics before reaching late-time saturation.
The ballistic-to-diffusive crossover time is $t_{1}^c = 2 \tau_m (\tau_a + \tau_m)/\tau_a$, as in case ($i$) due to $\tau_m \ll \tau_a, \tau_k$, while the diffusive-to-sub-diffusive crossover occurs at $t_{2}^c = \tau_k/\pi$, and the sub-diffusive-to-saturation crossover is $t_{3}^c = (\pi \tau_a^4)/[(\tau_a + \tau_m)(4 \tau_a^2 + \tau_a \tau_k + \tau_k \tau_m)]$, giving $t_1^c = 0.02$, $t_2^c = 1.59$, and $t_3^c = 77.56$ for this parameter set.

{\bf ($vi$) $( \t_k, \t_a)  \ll t \ll \t_m $ }: In this regime, ${\cal B}_v = \sqrt{\tau_k}/(2 \gamma^2 \sqrt{\tau_m})$ with $(1 + \omega^2 \tau_a^2) \approx 1$, indicating that MSCV has effectively reached saturation, as in case ($iv$). The line plot ($vi$) in Fig.~\ref{fig:msd_msv}($a$) corresponds to $(\tau_m, \tau_k, \tau_a) = (10, 0.01, 0.1)$; although $\tau_k \ll \tau_m$ allows a potential ballistic-to-diffusive crossover at $t_1^c = \sqrt{\tau_m \tau_k} (\tau_a + \tau_m)/\tau_a = 31.93$, this is pre-empted by the earlier ballistic-to-saturation crossover at $t_2^c = \left[ 2 \tau_a \tau_m \sqrt{\tau_k (\tau_a + \tau_m)} / \sqrt{4 \tau_a^2 + \tau_a \tau_k + \tau_k \tau_m} \right]^{1/2} = 1.30$, so only the ballistic-to-saturation transition is observed for this parameter set.

Table~\ref{table:regime} summarizes the key behaviors across these regimes.

{\bf Steady state:}
\begin{figure*}[t]
    \centering
    \includegraphics[scale=0.65]{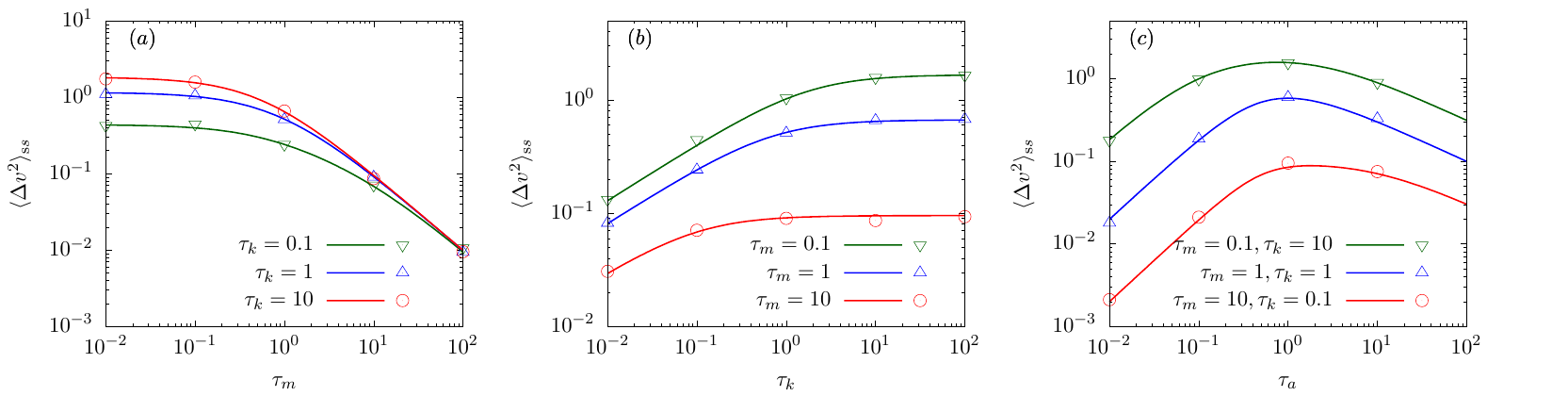}
    \caption{($a$) Variation of the steady-state  MSCV $\la \Delta v^2 \ra_{\rm ss}$ as a function of inertia $\t_m$ for a few values of $\t_k$ at $\t_a = 1$. ($b$) variation of $\D^v_{ss}$ as a function of interaction time $\t_k$ for a few values of $\t_m$ at $\t_a = 1$ and ($c$) variation of $\D^v_{ss}$ as a function of persistent time $\t_a$  for a few values of $\tau_{m}$ and $\tau_k$. The symbols denote the data points from the simulation, and the solid line is the plot of steady state MSCV presented in Eq.\eqref{eq:ss_mscv}. The activity was kept fixed as $v_0 = 1$.
    }
    \label{fig:msv_ss}
\end{figure*}
At long times, the MSCV reaches a steady-state value (see Appendix~\ref{sec:mscv_saturate}),
\begin{equation}
\label{eq:ss_mscv}
\langle {\D v}^2 \rangle_{\rm ss} = 
\frac{2 v_0^2 \tau_a \sqrt{\tau_k}}{\sqrt{(\tau_a + \tau_m)(4 \tau_a^2 + \tau_a \tau_k + \tau_m \tau_k)}}.
\end{equation}
Figures~\ref{fig:msv_ss}($a–c$) show the analytical prediction alongside simulation data, demonstrating excellent agreement. 
At small inertia ($\tau_m \to 0$), $\langle {\D v}^2 \rangle_{\rm ss}$ approaches 
$2 v_0^2 \sqrt{\tau_k/(4\tau_a + \tau_k)}$ and decreases as $1/\tau_m$ for large inertia [Fig.~\ref{fig:msv_ss}($a$)]. 
With increasing $\tau_k$, $\langle {\D v}^2 \rangle_{\rm ss}$ rises and saturates to 
$2 v_0^2 \tau_a/(\tau_a + \tau_m)$ [Fig.~\ref{fig:msv_ss}($b$)]. 
As a function of $\tau_a$, $\langle {\D v}^2 \rangle_{\rm ss}$ exhibits non-monotonic dependence--growing initially, reaching a maximum, and decaying as $1/\sqrt{\tau_a}$ at large persistence times  [Fig.~\ref{fig:msv_ss}($c$)].

The corresponding mean-squared velocity (MSV), $\langle v^2\rangle_{\rm ss}=\langle {\D v}^2 \rangle_{\rm ss}/2$, decreases with increasing inertia, increases with interaction timescale $\tau_k$, and shows a non-monotonic dependence on $\tau_a$, consistent with simulation results (Fig.~\ref{fig:msv_ss}). 
The steady-state kinetic temperature, defined as $k_B T_{\rm kin}=m\langle v^2\rangle_{\rm ss}$, thus reads
\[
k_B T_{\rm kin}=\frac{\gamma v_0^2\tau_a\tau_m\sqrt{\tau_k}}
{\sqrt{(\tau_a+\tau_m)(4\tau_a^2+\tau_a\tau_k+\tau_m\tau_k)}},
\]
increasing with inertia and saturating to $\gamma\tau_a v_0^2$ at large $\tau_m$. This defines an effective kinetic temperature that encodes both activity-induced fluctuations and interaction-mediated collective relaxation.

\begin{table*} 
    \centering
    \begin{tabular}{|c|c|c|c|c|c|c|}
    \hline
        Intermediate time $t$ & $\t_m \ll t \ll \t_a, \t_k$ & $\t_a \ll t \ll \t_m, \t_k$ & $\t_k \ll t \ll \t_m, \t_a$  & $\t_m, \t_a \ll t \ll \t_k$  & $\t_m, \t_k \ll t \ll \t_a$  & $\t_k, \t_a \ll t \ll \t_m$ \\
        \hline
        MSCV & $t$ & $t$ & $t$ & $t^0$ &  $t^{1/2}$ & $t^0$\\
        \hline
        MSD  & $t^2$ & $t^2$ & $t^2$ & $t$ & $t^2$ & $t$ \\
        \hline
    \end{tabular}
    \caption{Scaling of MSCV and MSD at intermediate time, as analyzed in Sec.~\ref {sec:second}.  }
    \label{table:regime}
\end{table*}

\subsubsection{Mean-squared displacement } \label{sec:msd}
We analyze the mean-squared displacement (MSD) of particles in the chain, which serves as a key measure for characterizing the influence of interactions within the system. The interplay between particle interactions, activity, and inertia can lead to anomalous transport behaviors -- such as subdiffusion and superdiffusion -- in addition to ordinary diffusion. In the following, we investigate possible crossovers in the MSD dynamics.

 The general two-time correlation between the displacement of the $l$-th and $m$-th particle is given by
\bea
{\cal C}^x_{l,m}(t) = \la [x_l(t) - x_l (0)][x_m (t) - x_m (0)] \ra = 2 C_{l,m}^x(0) - 2 C_{l,m}^x (t),
\eea
where the displacement correlation $C_{l,m}^x (t)$ is given by Eq.~(\ref{eq:corr_x}). 
Setting $l = m$ gives the MSD of the $l$-th particle as
\bea
    \label{eq:msd_lth}
  \langle {\Delta x}_l^2 \rangle (t) = {\cal C}^x_{l,l}(t) = \frac{ \gamma^2 v_0^2 \tau_a}{\pi} \int_{-\infty}^{\infty} \frac{\sum_k   \,\tilde{\mathcal{G}}_{l,k}(\omega)~ \tilde{\mathcal{G}}_{l,k}^{T}(-\omega)}{1+\omega^2 \tau_a^2} \left(2-2\cos \omega t\right)~ d\omega.
\eea
Under the assumption that all particles are statistically equivalent in the thermodynamic limit, the MSD of a tagged particle can be written as
\bea\label{bulk_msd_full_form}
	\langle {\Delta x}^2 \rangle (t) =\frac{1}{N}\sum_{l=1}^N {\cal C}^x_{l,l}(t) 
	=\frac{ \gamma^2 v_0^2 \tau_a}{ \pi} \int_{-\infty}^{\infty}  \mathcal{B}_x(\omega) \frac{(2-2 \cos \omega t)}{(1 +\omega^2 \tau_a^2)} ~d\omega\,,
\eea
where ${\cal B}_x (\omega)$ is given by (see Appendix \ref{appendix:bvomega_bxomega} )
\bea
    \label{eq:bx_omega}
    \mathcal{B}_x(\omega)=\frac{1}{N}\sum_{l,k} \tilde{\mathcal{G}}_{l,k}(\omega) \tilde{\mathcal{G}}_{k,l}(-\omega) 
    = \frac{1}{\gamma^2} {\mathrm{Re}} \left [  \frac{\sqrt{\tau_k}}{\omega^{3/2} \sqrt{( \omega \tau_m-i)(4 - \omega \tau_k (\omega \tau_m - i))}}\right].
 \eea
 
 An exact closed-form expression for $\la \Delta x^2 \ra (t)$ is difficult to obtain. 
However, numerical integration of Eq.~(\ref{bulk_msd_full_form}), together with the form of ${\cal B}_x(\omega)$, shows excellent agreement with the simulation results, as illustrated in Fig.~\ref{fig:msd_msv}($b$). 

To gain further insight, we analyze Eq.~(\ref{bulk_msd_full_form}) and employ a timescale-based approximation to characterize the MSD behavior across all relevant temporal regimes.

{\bf (A) Early time:} At short times, $t \ll (\tau_a, \tau_k)$, larger values of $\omega$ dominate, and $\mathcal{B}_x(\omega)$ can be approximated as 
$
\mathcal{B}_x(\omega) \approx \frac{1}{\gamma^2} {\mathrm{Re}} \left[ \frac{\sqrt{\tau_k}}{\omega^{3/2} \sqrt{- \omega \tau_k (\omega \tau_m - i)^2}} \right] 
= \frac{1}{\gamma^2} {\mathrm{Re}} \left[ \frac{1}{i\omega^{2} (\omega \tau_m - i)} \right] 
= \frac{1}{\gamma^2} \left[ \frac{1}{\omega^{2} (\omega^2 \tau_m^2 + 1)} \right].
$
This gives the MSD as
\begin{equation}
  \label{eq:msd_smalltime}
 \langle {\Delta x}^2 \rangle = \frac{v_0^2 \tau_a}{\pi} 
  \int_{-\infty}^{\infty} 
  \frac{4 \sin^2 (\omega t/2)}{\omega^{2} (\omega^2 \tau_m^2 + 1)(1+\omega^2 \tau_a^2)}\, d\omega 
  \approx \frac{v_0^2 \tau_a }{(\tau_a + \tau_m)} t^2 + \mathcal{O}(t^4).
\end{equation}
In this regime, the leading-order term corresponds to ballistic motion, where the amplitude depends on both the inertial and persistence times. This ballistic behavior is observed for all parameter values in Fig.~\ref{fig:msd_msv}($b$) at short times. 
{We note that this short-time result differs from Ref.~\cite{Marconi2024}, obtained for inertial AOUPs under the assumptions $t/\tau_a \ll 1$ and $\tau_a \ll \tau_m$.}

To facilitate the discussion of crossover times in different parameter regimes, we first examine the long-time behavior before addressing the intermediate-time regimes.

\vspace{0.3em}
{\bf (B) Long time:} In the long-time limit, $t \gg (\tau_a, \tau_k, \tau_m)$, the dominant contribution arises from small $\omega$, yielding 
$
{\cal B}_{x} (\omega) = \frac{\sqrt{\tau_k}}{2^{3/2} \gamma^2 \omega^{3/2}},
$
with $(1 + \omega^2 \tau_a^2) \approx 1$. The resulting MSD is
\begin{equation}
 \label{eq:msd_largetime}
\langle {\Delta x}^2 \rangle(t) = \frac{v_0^2 \tau_a \sqrt{\tau_k}}{ \pi } 
 \int_{-\infty}^{\infty} \frac{\sin^2(\omega t/2)}{\omega^{3/2}}\, d\omega 
 = 2 v_0^2 \tau_a \left( \frac{\tau_k}{\pi} \right)^{1/2} t^{1/2},
\end{equation}
which represents a single-file diffusion (SFD). This scaling resembles that observed in overdamped RTP chains~\cite{Paul24} and remains unaffected by inertia. 
{A similar late-time behavior was also reported in Ref.~\cite{Marconi2024} for chains of inertial AOUPs.}
The effective diffusion coefficient is given by $D_x = v_0^2 \tau_a \sqrt{\tau_k/\pi}$. This asymptotic behavior is consistent across all parameters shown in Fig.~\ref{fig:msd_msv}($b$).

\vspace{0.3em}
{\bf (C) Intermediate time regimes:}  
The MSD dynamics exhibit six distinct parameter regimes, analogous to those discussed for MSCV. We analyze each regime and identify the corresponding crossover times.

\noindent
{\bf ($i$) $\tau_m \ll t \ll (\tau_a, \tau_k)$:}  
Here, ${\cal B}_x \simeq 1/(\gamma^2 \omega^2)$, giving
\bea
\langle {\Delta x}^2 \rangle (t) = \frac{v_0^2 \tau_a}{\pi} \int_{-\infty}^{\infty} 
\frac{4 \sin^2(\omega t/2)}{\omega^2(1+\omega^2 \tau_a^2)}\, d\omega 
\approx v_0^2 t^2 + \mathcal{O}(t^3).
\eea
The MSD remains ballistic, crossing over to SFD at 
$t_{1}^c = [2 \tau_a (\tau_k/\pi)^{1/2}]^{2/3}$,
as obtained by equating this expression with Eq.~(\ref{eq:msd_largetime}).  
Parameters $(\tau_m, \tau_k, \tau_a) = (0.01, 10, 10)$ in the lineplot ($i$) in Fig.~\ref{fig:msd_msv}($b$) correspond to this regime, showing a crossover at $t_{1}^c = 10.84$.

\noindent
{\bf ($ii$) $\tau_a \ll t \ll (\tau_m, \tau_k)$:}  
In this case, ${\cal B}_x(\omega) \simeq \Re[1/(i\gamma^2 \omega^2(\omega\tau_m - i))] = 1/[\omega^2(\omega^2 \tau_m^2 + 1)]$, yielding
\bea
\langle {\Delta x}^2 \rangle (t) =  \frac{v_0^2 \tau_a}{\pi} 
\int_{-\infty}^{\infty} 
\frac{4 \sin^2(\omega t/2)}{\omega^2(1+\omega^2 \tau_m^2)}\, d\omega 
\approx \frac{v_0^2 \tau_a}{\tau_m} t^2 + \mathcal{O}(t^3).
\eea
The motion remains ballistic but now depends on inertia.  
The crossover to SFD occurs at $t_{1}^c = [2 \tau_m (\tau_k/\pi)^{1/2}]^{2/3}$.
For $(\tau_m, \tau_k, \tau_a) = (10, 100, 0.01)$, the lineplot ($ii$) in Fig.~\ref{fig:msd_msv}($b$) shows $t_{1}^c = 23.35$.

\noindent
{\bf ($iii$) $\tau_k \ll t \ll (\tau_m, \tau_a)$:}  
With ${\cal B}_x \simeq \sqrt{\tau_k/(4\tau_m)} \gamma^{-2}\omega^{-2}$,
\bea
\langle {\Delta x}^2 \rangle (t) = \frac{v_0^2 \tau_a}{2\pi} \sqrt{\frac{\tau_k}{\tau_m}}\, t^2 + \mathcal{O}(t^3),
\eea
indicating ballistic motion dependent on all three timescales.  
The crossover to SFD occurs at $t_{1}^c = [4 \pi (\tau_m/\pi)^{1/2}]^{2/3}$.  
Parameters $(\tau_m, \tau_k, \tau_a) = (10, 0.01, 10)$ yield $t_{1}^c = 7.95$ in the lineplot ($iii$) in Fig.~\ref{fig:msd_msv}($b$).

\noindent
{\bf ($iv$) $(\tau_m, \tau_a) \ll t \ll \tau_k$:}  
Here, ${\cal B}_x(\omega) \simeq 1/(\gamma^2 \omega^2)$ and $(1+\omega^2\tau_a^2) \approx 1$, giving
\bea
\langle {\Delta x}^2 \rangle(t) = \frac{v_0^2 \tau_a }{\pi} 
\int_{-\infty}^{\infty} \frac{4 \sin^2(\omega t/2)}{\omega^2}\, d\omega 
\approx 2 v_0^2 \tau_a t.
\eea
This regime shows normal diffusion with coefficient $D = v_0^2 \tau_a$, independent of inertia.  
The ballistic-diffusive and diffusive-SFD crossovers occur at $t_{1}^c = 2(\tau_a + \tau_m)$ and $t_{2}^c = \tau_k/\pi$, respectively.
For $(\tau_m, \tau_k, \tau_a) = (0.01, 100, 0.01)$, we obtain $t_{1}^c = 0.04$ and $t_{2}^c = 31.83$ in the lineplot ($iv$) in Fig.~\ref{fig:msd_msv}($b$).

\noindent
{\bf ($v$) $(\tau_m, \tau_k) \ll t \ll \tau_a$:}  
In this limit, ${\cal B}_x(\omega) = \sqrt{\tau_k}/(2\sqrt{2}\gamma^2 \omega^{3/2})$ with $(1+\omega^2\tau_a^2) \approx \omega^2\tau_a^2$, yielding
\bea
\langle {\Delta x}^2 \rangle (t) = \frac{v_0^2 }{2 } \sqrt{\frac{\tau_k}{\tau_a}}\, t^2 + \mathcal{O}(t^3).
\eea
The motion is ballistic, with a crossover to SFD at 
$t_{1}^c = [4 \tau_a (\tau_a/\pi)^{1/2}]^{2/3}$.  
For $(\tau_m, \tau_k, \tau_a) = (0.01, 5, 100)$, $t_{1}^c = 172.05$ in the lineplot ($v$) in Fig.~\ref{fig:msd_msv}($b$).

\noindent
{\bf ($vi$) $(\tau_k, \tau_a) \ll t \ll \tau_m$:}  
Here, ${\cal B}_x = \sqrt{\tau_k}/(2\gamma^2 \omega^2 \sqrt{\tau_m})$ with $(1+\omega^2\tau_a^2) \approx 1$, leading to
\bea
\langle {\Delta x}^2 \rangle (t) = \frac{v_0^2 \tau_a \sqrt{\tau_k}}{2\pi \sqrt{\tau_m}} 
\int_{-\infty}^{\infty} \frac{4 \sin^2(\omega t/2)}{\omega^2}\, d\omega 
\approx v_0^2 \tau_a \sqrt{\frac{\tau_k}{\tau_m}}\, t.
\eea
The motion is diffusive, with coefficient 
$D = \frac{v_0^2 \tau_a}{2}\sqrt{\frac{\tau_k}{\tau_m}}$.
Crossover times are 
$t_{1}^c = (\tau_a + \tau_m)\sqrt{\tau_k/\tau_m}$ (ballistic - diffusive) and 
$t_{2}^c = 4\tau_m/\pi$ (diffusive - SFD).  
For $(\tau_m, \tau_k, \tau_a) = (10, 0.01, 0.1)$, these correspond to 
$t_{1}^c = 0.32$ and $t_{2}^c = 12.73$ in the lineplot ($vi$) in Fig.~\ref{fig:msd_msv}($b$).

\vspace{0.3em}
We summarize the results for all intermediate regimes in Table~\ref{table:regime}. 
For a finite chain, overall diffusion emerges beyond the finite-size timescale $\tau_N \sim N^2 \tau_k$, where $N$ is the number of beads. In a chain with pinned boundaries, the dynamics of boundary particles differ from those in the bulk, and the MSD of bulk particles saturates beyond $\tau_N \sim N^2 \tau_k$. The case of a pinned boundary chain is discussed in Appendix~\ref{sec:pinned_boundary}, along with the evolution of MSCV and MSD.

\subsubsection{Spatio-temporal correlations}

\begin{figure}[t]
    \centering
    \includegraphics[width=0.8\linewidth]{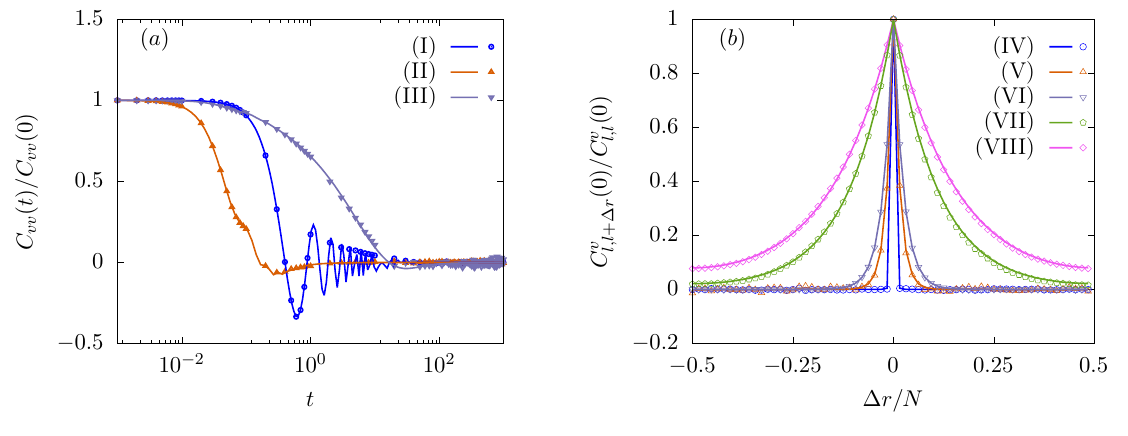}
    \caption{Velocity autocorrelation and spatial velocity correlations. 
($a$) Normalized velocity autocorrelation $C_{vv}(t)/C_{vv}(0)$ and ($b$) normalized equal-time spatial velocity correlation $C_{l,l+\Delta r}^v(0)/C_{l,l}^v(0)$ as a function of scaled distance $\Delta r/N = (m - N/2)/N$. 
Parameter sets $(\tau_m, \tau_k, \tau_a)$ are (I) (10, 0.01, 0.1), (II) (0.1, 0.01, 0.1), (III) (0.1, 0.01, 10), (IV) (0.1, 10, 0.1), (V) (0.1, 10, 10), (VI) (0.1, 0.02, 0.1), (VII) (10, 0.1, 10), and (VIII) (0.1, 0.1, 10), with fixed activity $v_0 = 1$. 
Oscillatory decay in ($a$) occurs at large inertia, while small inertia yields monotonic decay. 
In ($b$), the correlation length increases for weak interaction, small inertia, or large persistence. 
Points denote simulation results and solid lines represent numerical integration of Eq.~(\ref{eq:corrv_comp}). 
Simulations and integrations in ($b$) were performed with $N = 64$ beads, whereas panel ($a$) used $N=256$.
}
 \label{fig:time_space_corrv}
\end{figure}

The integral for finite $t$ in Eq.~(\ref{eq:autocorr}) does not admit a closed-form expression, so we evaluate it numerically. 
Fig.~\ref{fig:time_space_corrv}($a$) shows normalized velocity autocorrelation as a function of time for different parameters, compared with numerical simulations (symbols), showing good agreement. For the underdamped case (I), inertia $\tau_m$ and the restoring interaction $k$ induce oscillations as the velocity relaxes. For large persistence time $\tau_a$ (III), these effects are suppressed, yielding a smooth decay. When inertial and persistence times are comparable (II), the interplay of these timescales produces more complex behavior.

Fig.~\ref{fig:time_space_corrv}($b$) shows normalized equal-time spatial velocity correlation, $C_{l,l+\Delta r}^{v}(0)/C_{l,l}^{v}(0)$ using Eq.\eqref{eq:corrv_comp}, plotted against the scaled distance $\Delta r/N = (m-N/2)/N$ for various parameters, compared with simulations (symbols), showing good agreement~\cite{Caprini21}. For small inertia, persistence time, and interaction strength (IV), correlations are short-ranged and delta-like. Increasing $\tau_a$ (V) or interaction strength (VI) extends the correlation length. At small inertia, strong persistence and interactions yield long-range correlations (VIII), whereas larger inertia reduces them (VII). 

\section{Higher-Order Statistics and Probability Distributions}
The nonlinear nature of active Brownian particles (ABPs) leads to dynamics that differ markedly from the linear active 
Ornstein-Uhlenbeck particle (AOUP) model. These deviations are reflected in departures from Gaussian statistics, which we quantify using the excess kurtosis and full probability distributions over time. The ABP dynamics are governed by
\begin{align}
\dot{X} &= V, \nonumber\\
m \dot{V} + \gamma V &= -\Phi X + \gamma U, \nonumber\\
\dot{\Theta} &= \sqrt{\tfrac{2}{\tau_{\rm abp}}}\, \boldsymbol{\eta}(t),
\end{align}
where \(U = v_{\rm abp}(\cos\theta_1, \cos\theta_2, \ldots, \cos\theta_N)\) with $\Theta = (\theta_1, \theta_2,...,\theta_N)$ representing the orientation for ABPs, \(\Phi\) is the tridiagonal coupling matrix (Sec.~\ref{sec:model}), and \(\boldsymbol{\eta}\) denotes Gaussian white noise with \(\langle \eta_i(t)\eta_j(t') \rangle = \delta_{ij}\delta(t-t')\). The parameters are related to the generic active particle model through \(v_{\rm abp} = \sqrt{2}\,v_0\) and \(\tau_{\rm abp} = \tau_a\) (see Appendix~\ref{sec:mapping}).

\subsection{Excess kurtosis of velocity and displacement} \label{sec:fourth}
\begin{figure}[t]
    \centering
    \includegraphics[width=0.7\linewidth]{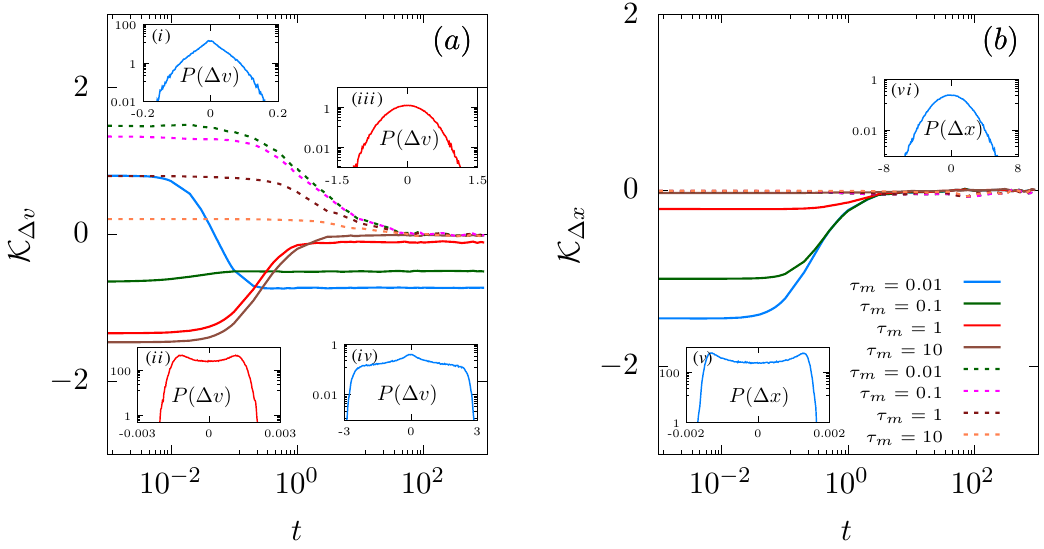}
    \caption{
Excess kurtosis for velocity and displacement, ${\cal K}_{\Delta v}$ ($a$) and ${\cal K}_{\Delta x}$ ($b$), versus time for inertia values given in the legend of panel ($b$). We use $v_{\rm abp} = \sqrt{2}$. 
Solid lines correspond to $\tau_{\rm abp} = 0.05$, $\tau_k = 20$, while dashed lines correspond to $\tau_{\rm abp} = 50$, $\tau_k = 1$. 
Insets ($i$) -- ($iv$) show the probability distributions of velocity changes, and insets ($v$) -- ($vi$) those of displacements.
Insets ($i$), ($iv$), ($v$), and ($vi$) correspond to the blue-curve parameters: $\tau_{\rm abp} = 0.05$, $\tau_k = 20$, $\tau_m = 0.01$, while ($ii$) and ($iii$) correspond to the red-curve parameters with $\tau_m = 1$ and identical remaining parameters. 
Insets ($i$), ($ii$), and ($v$) are evaluated at $t = 0.001$, and ($iii$), ($iv$), and ($vi$) at $t = 100$.
}
    \label{fig:kurx_kurv}
\end{figure}

We quantify deviations from Gaussian statistics by computing the excess kurtosis for a chain of inertial ABPs, defined as  
\begin{align}
\label{eq:kurtosis}
{\cal K}_{\Delta y}(t) = \frac{\langle \Delta y^4 \rangle}{\langle \Delta y^2 \rangle^2} - 3,
\end{align}
where \(\Delta y = y(t) - y(0)\) and \(y = x, v\) correspond to displacement and velocity, respectively. A zero value indicates a Gaussian distribution, while positive and negative values correspond to non-Gaussian departures. Figure~\ref{fig:kurx_kurv} shows the time evolution of \({\cal K}_{\Delta v}\) and \({\cal K}_{\Delta x}\) for different persistence times \(\tau_{\rm abp}\), interaction strengths \(\tau_k\), and inertias.

For velocity fluctuations [Fig.~\ref{fig:kurx_kurv}($a$)], \({\cal K}_{\Delta v}\) exhibits short-time positive and negative departures, corresponding to heavy-tailed and bimodal distributions (insets ($i$) and ($ii$)). These non-Gaussian features arise from activity and vanish for weak persistence. At long times, \({\cal K}_{\Delta v}\) approaches zero or slightly negative values, indicating Gaussian (inset ($iii$)) or finite-support unimodal distributions (inset ($iv$)). The latter occurs for small inertia and coupling (blue, green), while large inertia or strong coupling restores Gaussian behavior (red, brown).

The displacement kurtosis \({\cal K}_{\Delta x}\) [Fig.~\ref{fig:kurx_kurv}($b$)] is initially negative or zero, reflecting bimodal  (inset ($v$)) or Gaussian displacement statistics. At later times, \({\cal K}_{\Delta x}\) vanishes across all parameters, consistent with Gaussian steady-state behavior (inset ($vi$)).

\subsection{Probability distributions} \label{sec:prob}
In the previous section, we identified non-Gaussian deviations in the probability distributions, characterized by negative kurtosis corresponding to unimodal finite-support or bimodal distributions, and positive kurtosis associated with heavy-tailed distributions.

In this section, we quantitatively examine their evolution with respect to velocity and displacement increments, $\Delta v$ and $\Delta x$. Distinct data collapses are observed within different temporal scaling regimes identified from the evolution of the second moments, confirming the consistency and robustness of the underlying scaling behavior.

\begin{figure}[t]
    \centering
    \includegraphics[width=1\linewidth]{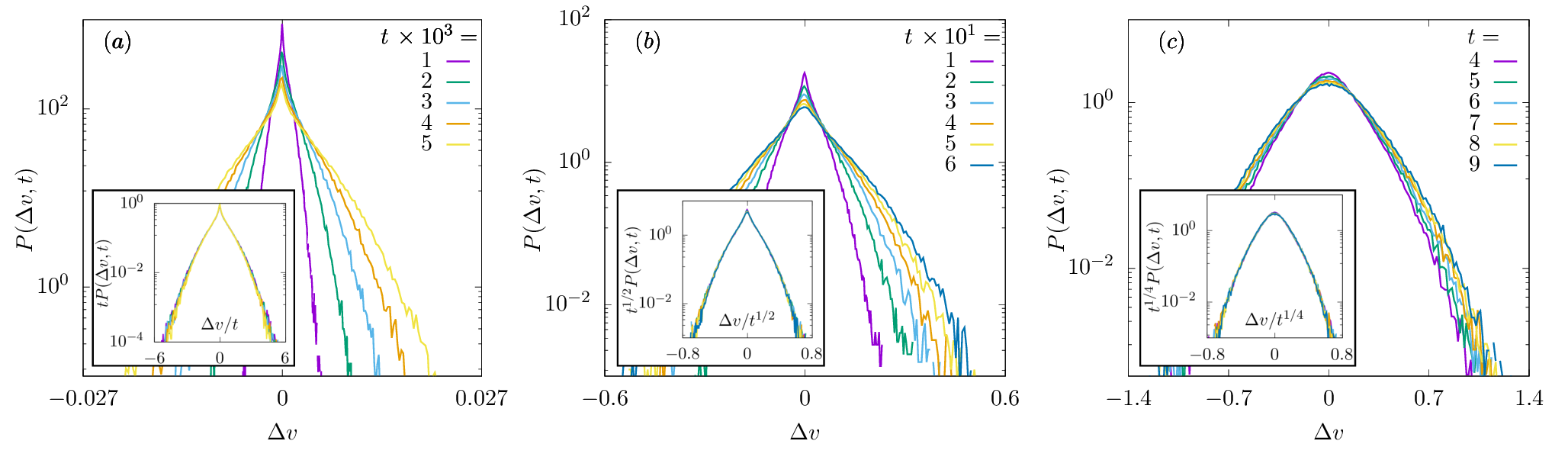}
    \caption{($a$)–($c$) Time evolution and scaling collapse of $P(\Delta v, t)$. Insets show data collapse with $P(\Delta v, t) = t^{-\mu} f_{\Delta v}(\Delta v/t^\mu)$: $\mu = 1$ (ballistic, $a$), $1/2$ (diffusive, $b$), $1/4$ (sub-diffusive, $c$).  Parameters $v_{abp}=\sqrt{2}$, $\t_m=0.01$, $\t_k=5$, $\t_a=100$ are the same as for line plot ($v$) in Fig.~\ref{fig:msd_msv}($a$) and ($b$).
 }
    \label{fig:fpv_data_collapse}
\end{figure}

In Fig.~\ref{fig:fpv_data_collapse}, we examine the time evolution of the velocity-change distribution $P(\Delta v, t)$ using parameters from the line plot ($v$) in Fig.~\ref{fig:msd_msv}($a$). The MSCV exhibits successive crossovers from ballistic to diffusive, sub-diffusive, and finally to steady-state saturation. At short times, $P(\Delta v, t)$ is unimodal with heavy tails and scales as $P(\Delta v, t) = t^{-1} f_1(\Delta v/t)$ [Fig.~\ref{fig:fpv_data_collapse}($a$)], though it can become bimodal for negative excess kurtosis (inset ($ii$) in Fig.~\ref{fig:kurx_kurv}($a$); see Appendix~\ref{sec:collapse_2}). Beyond $t_1^c = 2(\tau_a + \tau_m)/\tau_a$, the distribution follows diffusive scaling $P(\Delta v, t) = t^{-1/2} f_2(\Delta v/t^{1/2})$ [Fig.~\ref{fig:fpv_data_collapse}($b$)], and for $t > t_2^c,\, t_3^c$ becomes sub-diffusive with $P(\Delta v, t) = t^{-1/4} f_3(\Delta v/t^{1/4})$ [Fig.~\ref{fig:fpv_data_collapse}($c$)]. 
For the steady-state distributions of velocity and velocity-change, see Appendix~\ref{sec:pdv_dis_ss} and \ref{sec:pv_ss}.

\begin{figure}[t]
    \centering
    \includegraphics[width=1\linewidth]{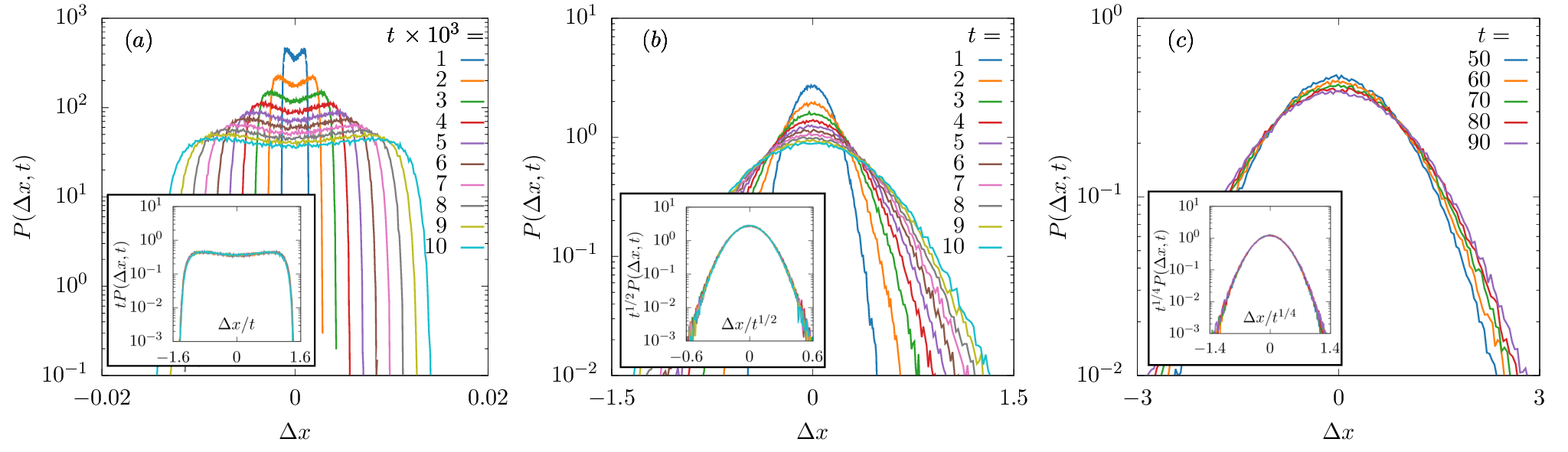}
     \caption{($a$)–($c$) Time evolution and scaling collapse of $P(\Delta x, t)$. Insets show data collapse with $P(\Delta x, t) = t^{-\mu} f_{\Delta x}(\Delta x/t^\mu)$ with $\mu = 1$ (ballistic, $a$), $1/2$ (diffusive, $b$), and $1/4$ (SFD, $c$). Parameters $v_{abp}=\sqrt{2}$, $\t_m=0.01$, $\t_k=100$, $\t_a=0.01$ are the same as for line plot ($iv$) in Fig.~\ref{fig:msd_msv}($a$).
}
    \label{fig:px_data_collapse}
\end{figure}

For the displacement distribution $P(\Delta x, t)$, we use parameters from line plot ($iv$) in Fig.~\ref{fig:msd_msv}($b$). The MSD shows a ballistic regime at short times, where $P(\Delta x, t)$ is bimodal with peaks at $\pm v_0 t$ due to activity, following the scaling $P(\Delta x, t) = t^{-1} f_4(\Delta x/t)$ [Fig.~\ref{fig:px_data_collapse}($a$)]. As the interplay of inertia and activity smooths these peaks, the distribution becomes unimodal. At intermediate times, the dynamics are diffusive with Gaussian scaling $P(\Delta x, t) = t^{-1/2} f_5(\Delta x/t^{1/2})$ [Fig.~\ref{fig:px_data_collapse}($b$)], and at late times, the system enters the single-file diffusion regime with $P(\Delta x, t) = t^{-1/4} f_6(\Delta x/t^{1/4})$ [Fig.~\ref{fig:px_data_collapse}($c$)]. The data collapse across regimes confirms the predicted scaling behavior discussed in Sec.~\ref{sec:second}.

\section{Discussion}
\label{sec:conclusion}
We developed a framework to investigate the dynamics of an inertial chain of active particles coupled via harmonic interactions in one dimension. Using an analytical Green's function approach, we derived the mean-squared change in velocity (MSCV) and mean-squared displacement (MSD), capturing all dynamical regimes governed by inertial ($\tau_m$), persistence ($\tau_a$), and interaction ($\tau_k$) timescales.

We identified six intermediate time regimes in MSCV and MSD, spanning ballistic, diffusive, and subdiffusive behavior, with MSCV saturating at steady state. The steady-state MSCV decreases with inertia and interaction and varies nonmonotonically with persistence, while MSD crossovers show inertia- or interaction-dominated scaling at intermediate times and subdiffusive behavior at late times. The crossover times and scaling coefficients are derived explicitly, and the second-moment results are consistent across ABP, RTP, and AOUP models.

Beyond analyzing second moments, we examined the full velocity and displacement distributions and the excess kurtosis for ABPs, uncovering pronounced non-Gaussian behaviors. These include kurtosis sign reversals and distributions that are bimodal, finite-support, or heavy-tailed, highlighting clear differences from AOUPs. Moreover, the time-dependent distributions display distinct data collapses in different temporal regimes, demonstrating the robustness of the underlying scaling behavior.

Together, these results show how inertia, activity, and interparticle coupling jointly shape multiscale relaxation in active chains. 
The framework links multiparticle interactions to microscopic dynamics, revealing experimentally accessible signatures of inertia in active matter with potential relevance for active granular particles and active solids. 
Our predictions can be directly tested in experimental setups as in Refs.~\cite{DolaiDasKunduDasguptaDharKumar2020, Khatri2012, Lin2016}, where one-dimensional chains of interacting granular particles are confined and tracked, allowing detailed measurements of particle displacements, velocities, and correlations to probe the predicted inertial and active dynamics.

\section*{Data availability}
All data generated or analysed during this study are included in the article. 

\section*{Author Contributions}
D.C. conceived and supervised the study. M.P. performed the calculations and drafted the manuscript with input from S.P.; D.C. finalized the manuscript.

\section*{Acknowledgements}
D.C. acknowledges a research grant from DAE (1603/2/2020/IoP/R\&D-II/15028), support from a Visiting Professorship at CY Cergy Paris Universit{\'e}, and an Associateship at ICTS-TIFR, Bangalore. S.P. thanks University of Delhi for financial assistance through the Faculty Research Programme under Grant-IOE (Ref. No. IOE/2024-25/12/FRP). S.P. also acknowledges IOP Bhubaneswar for hospitality.

\appendix
\section{Mapping across ABP, RTP and AOUP models}\label{sec:mapping}
The inertial Langevin dynamics of a free active particle in $1d$ reads
\bea
\dot x &=& v \nn\\
m \dot v &=& - \gamma v + \gamma v^a
\eea
where $v^a$ denotes the active velocity of the particle. 
For ABP, the active velocity $v^a = v_{\rm abp} \cos \theta$, where $\theta$ is particle orientation and evolves stochastically as $\dot \theta = \sqrt{2/\tau_{\rm abp}} \, \eta_r$.
For RTP, the active velocity $v^a = v_{\rm rtp} \sigma$, where $\sigma = \pm 1$ and follows the correlation $\la \sigma (t) \sigma (t') \ra = e^{-|t-t'|/\tau_{\rm rtp}}$.
For AOUP, active velocity $v^{a}$ follows $ \t_{\rm ao}\dot{v}^a =- v^a + v_{\rm ao} \sqrt{2\tau_{\rm ao}} \,\zeta$ with $\zeta$ being Gaussian white noise with unit deviate.
Thus, the active force in all three models is defined by $\la v^{a} \ra = 0$ and 
\begin{equation}
\left\langle v^a(t)\, v^a(t') \right\rangle =
\begin{cases}
\frac{v_{\rm abp}^{2}}{2}\, e^{-|t-t'|/\tau_{\rm abp}}, & \text{for ABP}, \\[6pt]
v_{\rm rtp}^{2}\, e^{- |t-t'|/\tau_{\rm rtp}}, & \text{for RTP}, \\[6pt]
v_{\rm ao}^{2}\, e^{-|t-t'|/\tau_{\rm ao}}, & \text{for AOUP}.
\end{cases}
\end{equation}
This leads to the mapping $v_{\rm abp}/\sqrt{2} = v_{\rm rtp} = v_{\rm ao}$ and $\tau_{\rm abp} = \tau_{\rm rtp} = \tau_{\rm ao}$.
The moment of dynamics agrees with each other up to the second moments. Note that following the persistence time and active force correlation, the ABP model can be mapped to a generic active particle considered in Section \ref{sec:model} and beyond.
The exact mapping is $v_{\rm abp} = \sqrt{2}\, v_0$ and $\t_{\rm abp} = \t_a$.

\section{Numerical integration}
\label{app_verlet}
The equations of motion were integrated using the standard velocity-Verlet algorithm. For each particle $i$, the positions, velocities, and orientations were updated as
\bea
x_{i} (t + \delta t) &=& x_i(t) + v_i (t) \delta t + \frac{\delta t^2}{2 \t_m} \left(f_{i}(t) -  v_i (t) \right) \\
v_i(t + \delta t) &=& \left(\frac{2 \t_m - \delta t}{2 \t_m +  \delta t} \right) v_i(t) + \left( \frac{\delta t}{2 \t_m +  \delta t} \right) \left( f_i(t)  + f_i(t+\d t) \right)\\
\theta_{i}(t+\delta t) &=& \theta_{i}(t) + (2/{\tau_{\rm abp}} )^{1/2}\,  dB_i (t) 
\eea
where $dB_i (t)$ is a Wiener process obeying $\la d B_i\ra =0$ and $\la d B_i(t) d B_j(0)\ra = \d_{ij} \,\d t$. The total force is given by
\bea
f_{i}(t) =  v_{\rm abp} \cos \theta_{i}(t) - \t_k^{-1} [2 x_i(t) - x_{i+1}(t) - x_{i-1}(t) \,] .
\eea
The integration time step $\delta t$ was chosen as $10^{-2} \times \min(\tau_m,\, \tau_k,\, \tau_a)$ or $10^{-4}$, whichever was smaller.
Simulations were performed until reaching steady state, before data collection.
For pinned boundary conditions (Appendix~\ref{sec:pinned_boundary}\,), the end-particle interactions were modified as in  Eq.~\ref{eq:pinned}.

\section{Calculation of $\tilde{V}(\omega)$} \label{appendix:vomega}
The Langevin equation for the velocity vector $V = (v_1, v_2,...,v_N)$ is given by
\bea
m \dot{V} + \gamma V = - \Phi X + \gamma V^a.
\eea
Using the Fourier transform in time, we obtain 
 \bea
 \tilde{V}_\omega = \frac{-\Phi \tilde{X}_\omega + \gamma  \tilde{V}^a_\omega}{i m \omega \mathds{I} + \gamma \mathds{I} },
 \eea
where $\tilde{X}_\omega = (1/2 \pi) \int_{- \infty}^{\infty} dt\, X(t)\, e^{-i \omega t} $, $\tilde{V}_\omega = (1/2 \pi) \int_{- \infty}^{\infty} dt\, V(t)\, e^{-i \omega t}$ and $\tilde{V}^a_\omega = (1/2 \pi) \int_{- \infty}^{\infty} dt\, V^a(t)\, e^{-i \omega t}$.
We use the expression of $\tilde{X}_\omega$ as in Eq.~(\ref{eq:xtilde_omega})
in the above expression to simplify it further to 
\bea
 \tilde{V}_\omega = \frac{\gamma } {i m \omega \mathds{I} + \gamma \mathds{I} } \left[ \frac{-\Phi}{\Phi + i \gamma \omega \mathds{I} -m \omega^2 \mathds{I}} + 1  \right] \tilde{V}^a_\omega  
 = \frac{i \gamma \omega \tilde{V}^a_\omega}{\Phi + i \gamma \omega \mathds{I} - m \omega^2 \mathds{I}} 
 = i \gamma \omega \tilde{{\cal G}}(\omega) \tilde{V}^a (\omega).
\eea
We use the above expression in Eq.~(\ref{eq:generic_corrv}) to get the general form of two-time velocity correlation and MSCV.

\section{Calculation of $\mathcal{B}_v(\omega)$ and $\mathcal{B}_x(\omega)$} \label{appendix:bvomega_bxomega}
Here, we provide the explicit calculation of ${\cal B}_x (\omega)$, as given in Eq.~(\ref{eq:bx_omega}), and its extension to obtain ${\cal B}_v(\omega)$, as shown in Eq.~(\ref{eq:bvomega}).
Noting that $\Phi$ is a tridiagonal matrix, whose eigenvalues can be expressed as
\begin{equation}
\label{eq:eigen_value}
	\lambda_{q}=2k(1-\cos q)\,, ~~~~q=\frac{s \pi}{N+1}, ~~s=1,2,\dots N\,,
\end{equation}
 we rewrite Eq.~(\ref{eq:bx_omega}) as
\begin{eqnarray}
    \label{eq:qintegral}\mathcal{B}_x(\omega)&=&\frac{1}{N}\sum_{l,k} \tilde{\mathcal{G}}_{l,k}(\omega) \tilde{\mathcal{G}}_{k,l}(-\omega) =\frac{1}{N} \sum_q \frac{1}{(\lambda_{q}-m\omega^2)^2+ \gamma^2 \omega^2}\nonumber \\
	&=&\frac{1}{2\pi}\int_{0}^{2\pi} \frac{dq}{4k^2(1-\cos q)^2-4k(1-\cos q)m\omega^2+m^2\omega^4+\omega^2\gamma^2}\,.
\end{eqnarray}

We define a unit circle $z = e^{iq}$ and rewrite the above expression as
\begin{equation}
    \mathcal{B}_x(\omega)= \frac{1}{2\pi} \oint \frac{-iz dz}{k^2(z-1)^4+2kz(z-1)^2m\omega^2+z^2(m^2\omega^4+\gamma^2\omega^2)}.
\end{equation}

In terms of the time scales $\tau_k = \gamma/k$ and $\tau_m=m/\gamma$, the integral takes the following form
\begin{align}
  \mathcal{B}_x(\omega) = \frac{-i\tau_k^2}{2 \pi \gamma^2} \oint \frac{z dz }{(z-1)^4 + 2 \tau_k \tau_m \omega^2 z(z-1)^2 + z^2 \tau_k^2(\omega^4 \tau_m^2 + \omega^2)}. 
  \label{eq:bw1}
\end{align}
The contour integral above has four poles, determined as the solutions of the equation
$(z-1)^4+2z(z-1)^2 \tau_m \tau_k \omega^2+\omega^2 z^2 \tau_k^2(\tau_m^2\omega^2+1)=0$.
These roots are in pairs of complex conjugates and are given by
\begin{align}
    &z_1 = 1 + \frac{1}{2} \left[ (i \omega \tau_k - \tau_k \tau_m \omega^2) - \sqrt{\omega \tau_k (\omega \tau_m - i)(-4-i \omega \tau_k + \tau_k \tau_m \omega^2)} \right] \\
    & z_1^* = 1 + \frac{1}{2} \left[ (-i \omega \tau_k - \tau_k \tau_m \omega^2) - \sqrt{\omega \tau_k (\omega \tau_m + i)(-4+i \omega \tau_k + \tau_k \tau_m \omega^2)} \right] \\
    & z_2 = 1 + \frac{1}{2} \left[ (i \omega \tau_k - \tau_k \tau_m \omega^2) + \sqrt{\omega \tau_k (\omega \tau_m - i)(-4-i \omega \tau_k + \tau_k \tau_m \omega^2)} \right] \\
     & z_2^* = 1 + \frac{1}{2} \left[ (-i \omega \tau_k - \tau_k \tau_m \omega^2) + \sqrt{\omega \tau_k (\omega \tau_m + i)(-4+i \omega \tau_k + \tau_k \tau_m \omega^2)} \right]
\end{align}

We analyze these roots by plotting $|z_1|^2$ and $|z_2|^2$ as a function of $\omega$ for different parameter values of $\t_k$, $\t_m$ and $\t_a$ and identify that $|z_2|^2$ increases from $1$ towards $\infty$ and $|z_1|^2$ decreases from $1$ towards $0$, as $\omega$ goes from $0$ to $\pm \infty$. Thus, for the unit circle, only the residues corresponding to $z_1$ and $z_1^*$ contribute to the integral, which yields
\begin{align}
    \mathcal{B}_x(\omega)&= \frac{-i\tau_k^2}{2 \pi \gamma^2} ~2\pi i [\text{Res}_{z \to z_1}+ \text{Res}_{z \to z_1^*}]= \frac{\tau_k^2}{ \gamma^2} \left[ \frac{z_1}{(z_1-z_1^*)(z_1-z_2)(z_1-z_2^*)}+ \frac{z_1^*}{(z_1^*-z_1)(z_1^*-z_2)(z_1^*-z_2^*)} \right] \nonumber\\
    &=\frac{\tau_k^2}{ \gamma^2} \left[ \frac{1}{2 \t_k^{3/2} \omega^{3/2} \sqrt{(\omega \t_m - i) (4 - \omega \t_k (\omega \t_m - i))}}  + \frac{-1}{2 \t_k^{3/2} \omega^{3/2} \sqrt{(\omega \t_m + i) (4 - \omega \t_k (\omega \t_m + i))}}\right].
\end{align}
As it turns out, the above quantity is real and can be expressed in a more compact form as
\begin{align}
    \mathcal{B}_x(\omega) &= \frac{1}{\gamma^2} {\mathrm{Re}} \left [  \frac{\sqrt{\tau_k}}{\omega^{3/2} \sqrt{( \omega \tau_m-i)(4 - \omega \tau_k (\omega \tau_m - i))}}\right].
\end{align}

\begin{figure}
    \centering
    \includegraphics[width=0.8\linewidth]{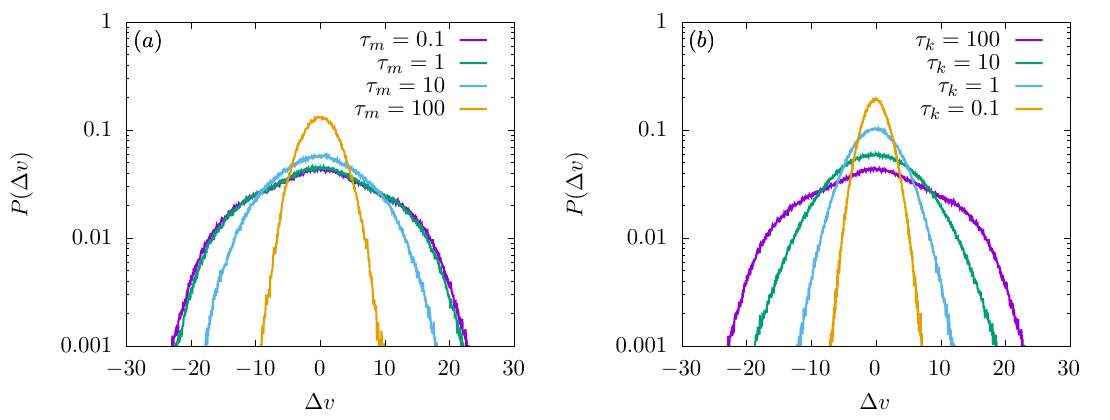}
    \caption{The late time probability distribution of change in velocity $P({\D v})$ as a function of inertia ($a$) and interaction time scale ($b$). The parameter values were taken as $\t_k = 100$ in ($a$) and $\tau_m = 0.1$ along with $\t_{\rm abp} = \t_a = 10$, and $v_{\rm abp} = \sqrt{2} \,v_0 = 10$ in both. } 
    \label{fig:pdv_ss}
\end{figure}

To calculate ${\cal B}_v(\omega)$, we follow the same steps to write 
\begin{align}
  \mathcal{B}_v(\omega) = \frac{-i\tau_k^2}{2 \pi \gamma^2} \oint \frac{\omega^2 z dz }{(z-1)^4 + 2 \tau_k \tau_m \omega^2 z(z-1)^2 + z^2 \tau_k^2(\omega^4 \tau_m^2 + \omega^2)}. 
\end{align}
The above contour integral has the same poles as ${\cal B}_x (\omega)$, and by following the same steps used for ${\cal B}_x (\omega)$, one can directly write
\bea
\mathcal{B}_v(\omega) = \frac{-i\tau_k^2 \omega^2}{2 \pi \gamma^2} ~2\pi i [\text{Res}_{z \to z_1}+ \text{Res}_{z \to z_1^*}] 
 = \frac{1}{\gamma^2} {\mathrm{Re}} \left [  \frac{\sqrt{\omega \tau_k}}{\sqrt{( \omega \tau_m-i)(4 - \omega \tau_k (\omega \tau_m - i))}}\right].
\eea
\section{MSCV and its steady-state limit}
\label{sec:mscv_saturate}
The MSCV can be expressed as 
\bea
\langle {\D v}^2 \rangle(t) &=& \frac{1}{N} \sum_{l = 1}^{N}{\cal C}^v_{l,l}(t)=\frac{\gamma^2 v_0^2 \t_a}{\pi} \int_{-\infty}^{\infty} \frac{(1/N) \sum_{l,k}\omega^2 \tilde{{\cal G}}_{l,k} (\omega) \tilde{{\cal G}}^T_{l,k} (-\omega) }{1 + \omega^2 \t_a^2} (2 - 2 e^{i \omega t}) d\omega \nn\\
&& =\frac{\gamma^2 v_0^2 \t_a}{\pi} \int_{-\infty}^{\infty} \frac{1}{N} \sum_q \frac{\omega^2}{[(\lambda_{q}-m\omega^2)^2+ \gamma^2 \omega^2]} \frac{(2- 2 e^{i \omega t})}{1 + \omega^2 \t_a^2} d\omega,
\eea
where $\lambda_q$ is the eigenvalue of the tridiagonal matrix as defined in Eq.~\ref{eq:eigen_value}. As $\lambda_q$ is periodic over $0$ to $2 \pi$, the above equation can be expressed as 
\bea
\langle {\D v}^2 \rangle(t) =\frac{\gamma^2 v_0^2 \t_a}{\pi} \int_{-\infty}^{\infty} \frac{dq}{2 \pi} \int_{0}^{2 \pi} \frac{\omega^2}{[(\lambda_{q}-m\omega^2)^2+ \gamma^2 \omega^2]} \frac{(2- 2 e^{i \omega t})}{1 + \omega^2 \t_a^2} d\omega.
\eea
At late time $t \to \infty$, noting that all dissipation terms vanish, we perform the $\omega$-integral first to obtain the late-time MSCV as
\bea
\label{eq:mscv_sat}
\langle {\D v}^2 \rangle_{\rm ss}&=&\frac{1}{2 \pi}\int_{0}^{2 \pi} \langle {\D v}^2 \rangle_q|_{\rm t \to \infty}\, dq  = \frac{1}{2 \pi} \int_{0}^{2 \pi}  \frac{2 v_0^2 \t_k \t_a}{2 \t_a^2(1 - \cos q) +   \t_k \t_a + \t_m \t_k} \,dq\nn\\
&&=  \frac{2 v_0^2 \t_a \sqrt{\t_k}}{\sqrt{(\t_a + \t_m)(4 \t_a^2 + \t_a \t_k + \t_m \t_k)}}.
\eea
At long times, the MSCV saturates, reaching the above steady-state form determined by inertia, activity, and interaction timescales.

\begin{figure}[t]
    \centering
    \includegraphics[width=0.8\linewidth]{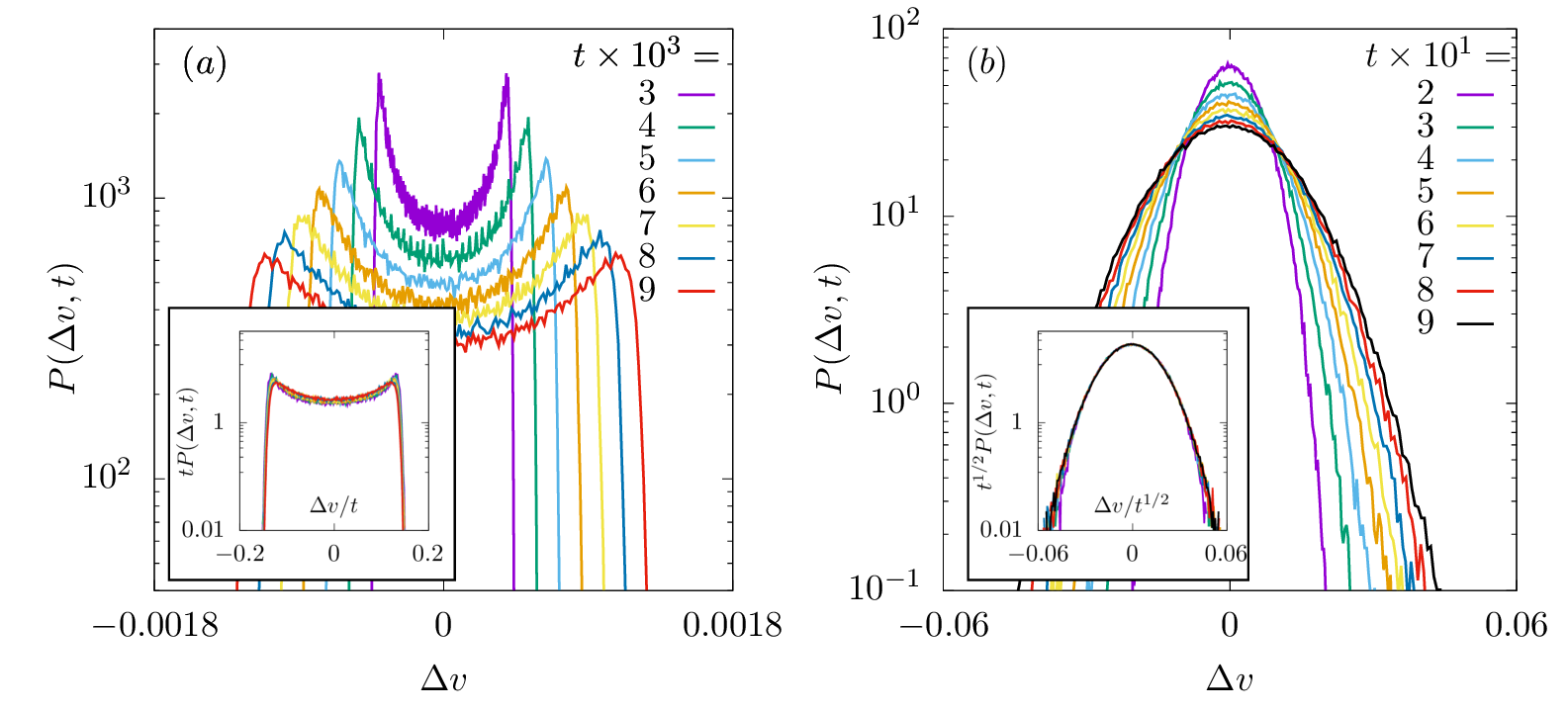}
    \caption{($a$–$b$)  Time evolution of the velocity-change distribution $P({\D v}, t)$ showing the scaling $P({\D v}, t)=t^{-\mu}f_{{\D v}}({\D v}/t^{\mu})$, with $\mu=1$ (ballistic, $a$) and $\mu=1/2$ (diffusive, $b$). Insets show the data collapse of the distributions. Parameters correspond to line plot ($ii$) in Fig.~\ref{fig:msd_msv}($a$).}
    \label{fig:pv_data_collapse}
\end{figure}

\section{Data collapse of probability distribution for change in velocity}
\label{sec:collapse_2}

Figure~\ref{fig:pv_data_collapse} shows the data collapse of the velocity-change distribution at the parameter value of line plot ($ii$) in Fig.~\ref{fig:msd_msv}($a$). The MSCV dynamics transition from ballistic to diffusive behavior before saturating at late times. In the ballistic regime, the bimodal ${\D v}$ distribution with peaks near $\pm v_0 t$ follows the scaling $P({\D v},t)=t^{-1}f_1({\D v}/t)$ [Fig.~\ref{fig:pv_data_collapse}($a$)], while in the diffusive regime it follows $P({\D v},t)=t^{-1/2}f_2({\D v}/t^{1/2})$ [Fig.~\ref{fig:pv_data_collapse}($b$)].

\section{Steady state distribution of velocity increment} \label{sec:pdv_dis_ss}
The steady-state probability distribution of velocity increments, $P(\Delta v)$, shows a crossover from a strongly non-Gaussian form at low inertia and weak interparticle interactions to a Gaussian form at high inertia and strong interactions.
 For small inertia or weak interactions (large $\tau_k$), activity produces a finite-support unimodal distribution (Fig.~\ref{fig:pdv_ss}). Increasing inertia at weak interaction strength drives a transition from this finite-support form to a Gaussian profile (Fig.~\ref{fig:pdv_ss}$a$). Similarly, increasing interaction strength (reducing $\tau_k$) at small inertia induces the same transition to Gaussian (Fig.~\ref{fig:pdv_ss}$b$).

\section{Steady state distribution of velocity} \label{sec:pv_ss}
\begin{figure}
    \centering
    \includegraphics[width=0.8\linewidth]{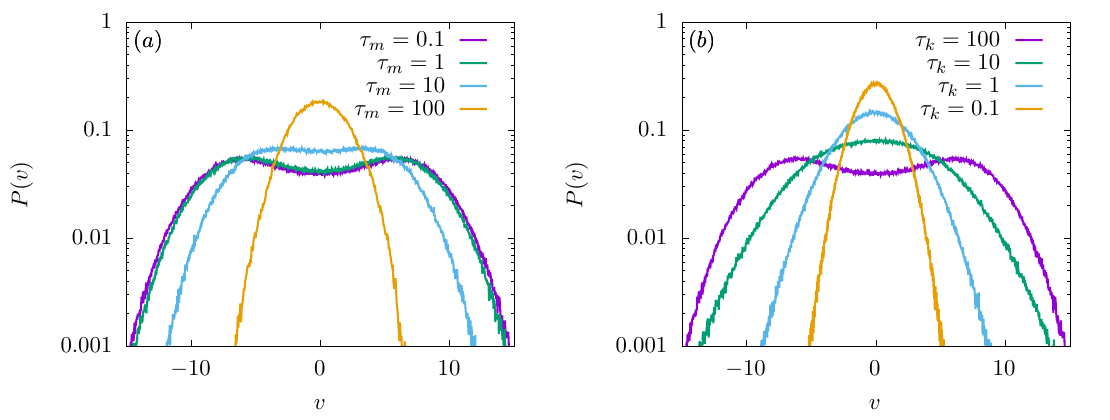}
    \caption{Steady-state velocity distribution, $P(v)$, as a function of inertia (a) and interaction timescale (b). Parameters are $\tau_k = 100$ in (a) and $\tau_m = 0.1$ in (b); $\tau_{\rm abp} = \tau_a = 10$, $v_{\rm abp} = \sqrt{2}\, v_0 = 10$ in both panels.} 
    \label{fig:pv_ss}
\end{figure}

The steady-state distribution $P(v)$ shows similar qualitative changes from non-Gaussian to Gaussian as for $P(\D v)$. For small inertia and weak interactions (large $\tau_k$), activity dominates, producing a bimodal $P(v)$ [Fig.~\ref{fig:pv_ss}($a$), $\tau_m = 0.1$]. Increasing inertia suppresses this effect, yielding a Gaussian distribution ($\tau_m = 100$), consistent with free ABP observations~\cite{Patel_2023}. Similarly, stronger interactions (smaller $\tau_k$) at low inertia drive a transition from bimodal to Gaussian [Fig.~\ref{fig:pv_ss}($b$)], reflecting the thermalizing role of interactions, as also reported in ABPs under a harmonic trap~\cite{Patel_2024}.

\section{MSD and MSCV of any tagged particle in boundary pinned chain} \label{sec:pinned_boundary}
 \begin{figure}[t]
    \centering
    \hskip -0.5 cm
    \includegraphics[width=17 cm]{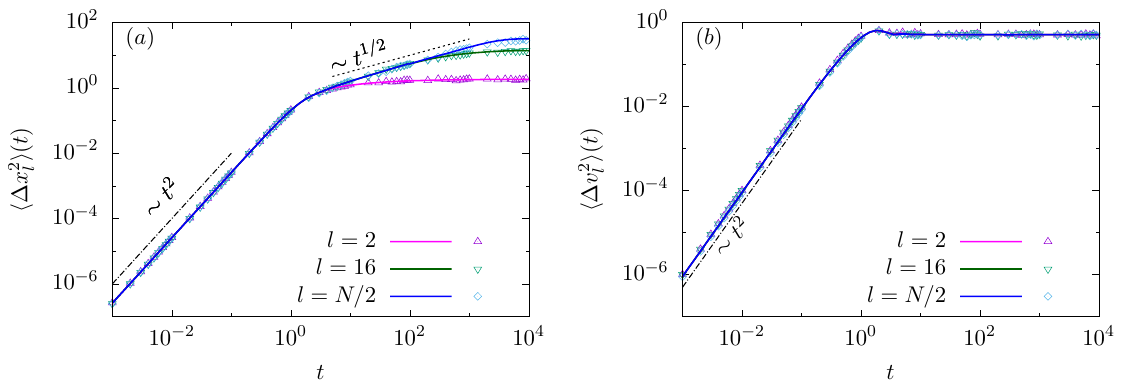}
    \caption{Plot of MSD ($a$) and MSCV ($b$) for second($l =2$), 16-th ($l = 16$) and the middle ($l = N/2$) particle for a chain of $N = 128$ particles with $\t_a = 0.5$, $\t_k=1.0$, $\t_m=1.0$. The dashed line corresponds to the power law behaviors showing ballistic $t^2$, and the SFD $t^{1/2}$ scaling. The solid lines are obtained by the numerical integration of Eq.~(\ref{msd_diag_matrix_2}) in ($a$) and Eq.~(\ref{tagged_msv_2}) in ($b$). }
    \label{fig:pinned_msd_msv} 
\end{figure}
 
For a pinned chain, we use external harmonic traps to constrain the dynamics of the two end particles of the chain. All other particles follow the same Langevin dynamics as given in Eq.~(\ref{eq:xdot}); however, the dynamics of the pinned boundaries ($i=1,\, N$) follow 
\bea
\label{eq:pinned}
\dot{x}_1 &=& v_1, \nn\\
m \dot{v}_1 &=& - \gamma v_1 - k (x_1 - x_2) - k x_1 + \gamma v^a_1, \nn\\
\dot{x}_N &=& v_N, \nn\\
m \dot{v}_N &=& - \gamma v_N - k (x_N - x_{N-1}) - k x_N + \gamma v^a_N. \nn\\
\eea

We find that the MSCV of individual tagged particles are invariant with respect to the boundary condition. The MSCV for each tagged particle is given by
\begin{align}\label{tagged_msv_2}
\langle {\D v_l}^2 \rangle (t) = \frac{ \gamma^2 v_0^2 \t_a }{\pi } \int_{-\infty}^{\infty} \frac{\sum_k  \omega^2 \,\tilde{\mathcal{G}}_{l,k}(\omega)~ \tilde{\mathcal{G}}_{l,k}^{T}(-\omega)}{1+\omega^2 \t_a^2} \left(2-2\cos \omega t\right)~ d\omega.
\end{align}\,
 Figure~\ref{fig:pinned_msd_msv}($b$) shows the MSCV for various tagged particles $l$, with continuous lines representing the numerical integration of Eq.~\eqref{tagged_msv_2}. All tagged particles exhibit MSCV behavior similar to that of bulk particles, displaying a crossover from ballistic motion to late-time saturation. 
 
 In contrast, under pinned boundary conditions, the MSD of boundary particles deviates from that of the bulk, as they are effectively confined in a local harmonic potential. Consequently, the tagged-particle dynamics evolve from the boundary toward the bulk. The MSD of the $l$-th tagged particle is given by Eq.~(\ref{eq:msd_lth}),
\begin{equation}\label{msd_diag_matrix_2}
\langle {\D x_l}^2 \rangle(t)={\mathcal{C}}_{l,l}^x(t)=\frac{ \gamma^2 v_0^2 \t_a}{\pi} \int_{-\infty}^{\infty} \frac{\sum_k \tilde{\mathcal{G}}_{l,k}(\omega)~ \tilde{\mathcal{G}}_{l,k}^{T}(-\omega)}{1+\omega^2 \t_a^2} \left(2-2\cos \omega t\right)~ d\omega\,. 
\end{equation}

The numerical integration of the above equation yields the evolution of tagged particles, illustrated for selected cases in Fig.~\ref{fig:pinned_msd_msv}($a$). Owing to the left–right symmetry about the central particle ($l = N/2$), the local behavior of the entire chain can be characterized by computing $\Delta x_l$ on one side only. As shown in Fig.~\ref{fig:pinned_msd_msv}($a$), boundary particles experience the pinning effect earlier and saturate before exhibiting SFD, whereas bulk particles ($l \sim N/2$) display clear SFD behavior prior to saturation.

\end{document}